\documentclass{article} 
\usepackage{nips15submit_e,times}
\usepackage{hyperref}
\usepackage{url}
\usepackage{color}
\usepackage{amsmath,amssymb}
\usepackage{algorithm}
\usepackage{algorithmic}
\usepackage{graphicx}
\usepackage{hyperref}
\usepackage{subfigure} 
\usepackage{relsize} 
\usepackage{array}
\usepackage{wrapfig}
\usepackage{multicol}

\nipsfinalcopy 
\title{Planar Ultrametric Rounding for Image Segmentation
\thanks{ JY acknowledges the support of Experian, CF acknowledges support of 
NSF grants IIS-1253538 and DBI-1262547 
} 
}

\author{
Julian Yarkony \\
Experian Data Lab\\
San Diego, CA 92130 \\
\texttt{julian.yarkony@experian.com} \\
\And
Charless C. Fowlkes \\
Department of Computer Science \\
University of California Irvine \\
\texttt{fowlkes@ics.uci.edu} \\
}

\newcommand{\cycpoly}{\mathsf{CYC}}              
\newcommand{\cyccone}{\cycpoly^\triangle}    

\newcommand{\mcut}{\mathsf{MCUT}}  
\newcommand{\mcutcone}{\mcut^\triangle} 

\newcommand{\cut}{\mathsf{CUT}}   
\newcommand{\cutcone}{\cut^\triangle} 

\begin{document}
\maketitle
\begin{abstract}
We study the problem of hierarchical clustering on planar graphs.  We formulate
this in terms of an LP relaxation of ultrametric rounding.  To solve this LP
efficiently we introduce a dual cutting plane scheme that uses minimum
cost perfect matching as a subroutine in order to efficiently explore the space
of planar partitions.  We apply our algorithm to the problem of hierarchical 
image segmentation.
\end{abstract}
\section{Introduction}
\label{sec:introduction}

In this work, we formulate hierarchical image segmentation from the perspective
of estimating an ultrametric over the set of image pixels that agrees closely
with an input set of noisy pairwise distances.  An ultrametric is a metric
space in which the triangle inequality is replaced by the ultrametric
inequality $d(u,v) \leq \max\{d(u,w),d(v,w)\}$.  This inequality captures the
transitive property of clustering (if $u$ and $w$ are in the same cluster and
$v$ and $w$ are in the same cluster, then $u$ and $v$ must also be in the same
cluster).  Thresholding an ultrametric immediately yields a partition into
sets whose diameter is less than the given threshold and varying the threshold
naturally produces a hierarchical clustering in which clusters at high
thresholds are composed of clusters at lower thresholds.

Inspired by the approach of \cite{ailon2005fitting}, our method represents an
ultrametric explicitly as a hierarchical collection of segmentations.
Determining the appropriate segmentation at a single distance threshold is
equivalent to finding a minimum-weight multicut in a graph with both positive
and negative edge weights \cite{Andres2012,highcc,ilpalg,cutgluecut,PlanarCC,
zhanyarko,AndresYarkony2013,baldiyark,bagon}. Finding an ultrametric imposes
the additional constraint that these multicuts are hierarchically consistent
across different thresholds.  We focus on the case where the input distances
are specified by a planar graph which arises naturally in the domain of image
segmentation where elements are pixels or superpixels and distances are defined
between neighbors.  This allows us to exploit fast combinatorial algorithms for
partitioning planar graphs that yield tighter LP relaxations than the local
polytope relaxation \cite{PlanarCC}. 

This paper is organized as follows.  We first introduce the ultrametric
rounding problem and the relation between multicuts and ultrametrics.  We then
introduce a LP relaxation that uses a delayed column generation approach that
exploits planarity to efficiently find cuts using the classic reduction to
minimum-weight perfect matching \cite{fisher2,Bar1,Bar2,Bar3}. We apply our
algorithm to the task of natural image segmentation on the Berkeley
Segmentation Data Set benchmark \cite{bsdspaper}.  We show compelling visual
results and demonstrate that our algorithm converges rapidly and produces  near
optimal or optimal solutions in practice with guarantees. 

\section{Ultrametric Rounding and Multicuts}

Let $G=(V,E)$ be a weighted graph with non-negative edge weights $\theta$
indexed by edges $e=(u,v) \in E$.  Our goal is to find an ultrametric distance
$d_{(u,v)}$ over vertices of the graph that is close to $\theta$ in the sense that
the distortion $\sum_{(u,v) \in E} \|\theta_{(u,v)} - d_{(u,v)}\|^2_2$ is
minimized.  We begin by reformulating this rounding problem in terms of finding a
set of nested multicuts in a family of weighted graphs.

We specify a partitioning or multicut of the vertices of the graph $G$ into
components using a binary vector $\bar{X} \in \{ 0,1 \}^{|E|}$ where
$\bar{X}_e=1$ indicates that the edge $e=(u,v)$ is ``cut'' and that the
vertices $u$ and $v$ associated with the edge are in separate components of the
partition.  We use $\mcut(G)$ to denote the set of binary indicator vectors
$\bar{X}$ that represent valid multicuts of the graph $G$. For notational
simplicity, in the remainder of the paper we frequently omit the dependence on
$G$ which is given as a fixed input.

A necessary and sufficient condition for an indicator vector $\bar{X}$ to
define a valid multicut in $G$ is that for every cycle of edges, if one edge on
the cycle is cut then at least one other edge in the cycle must also be cut.
Let $C$ denote the set of all cycles in $G$ where each cycle $c \in C$ is a set
of edges and $c-\hat{e}$ is the set of edges in cycle $c$ excluding edge
$\hat{e}$. We can express $\mcut$ in terms of these {\em cycle inequalities} 
as:
\begin{align}
\mcut = \left\{\bar{X} \in \{0,1\}^{|E|} : \sum_{e \in c -\hat{e}} \bar{X}_e \geq \bar{X}_{\hat{e}}, \forall \! c \in C, \hat{e}\in c \right\}
\label{cycineq}
\end{align}

A hierarchical clustering of a graph can be described by a nested collection of
multicuts.  We denote the space of valid hierarchical partitions with $L$
layers by $\bar{\Omega}_L$ which we represent by a set of $L$ edge-indicator vectors
$\mathcal{X}=(\bar{X}^1,\bar{X}^2,\bar{X}^3,\ldots,\bar{X}^L)$ in which any cut
edge remains cut at all finer layers of the hierarchy.
\begin{equation}
\label{ver0a}
\bar{\Omega}_L = \{ (\bar{X}^1,\bar{X}^2,\ldots \bar{X}^L ) : \bar{X}^l \in \mcut, \bar{X}^l \geq \bar{X}^{l+1} \; \forall l\}
\end{equation}

Given a valid hierarchical clustering $\mathcal{X}$, an ultrametric $d$ can be
specified over the vertices of the graph by choosing a sequence of real values
${0 = \delta^0 <\delta^1<\delta^2<\ldots<\delta^L}$ that indicate a distance
threshold associated with each level $l$ of the hierarchical clustering.  The
ultrametric distance $d$ specified by the pair $(\mathcal{X},\delta)$ assigns a
distance to each pair of vertices $d_{(u,v)}$ based on the coarsest level of the
clustering at which they remain in separate clusters. For pairs corresponding
to an edge in the graph $(u,v) = e \in E$ we can write this explicitly in terms
of the multicut indicator vectors as:
\begin{equation}
d_e = \max_{l \in \{0,1,\ldots,L\} } \delta^l \bar{X}^l_e = \sum^{L}_{l=0} \delta^l [\bar{X}_{e}^l > \bar{X}_{e}^{l+1}]\\
\end{equation}
We assume by convention that $\bar{X}_e^{0} = 1$ and
$\bar{X}_e^{L+1} = 0$.  Pairs $(u,v)$ that do not correspond to an edge in the
original graph can still be assigned a unique distance based on the coarsest
level $l$ at which they lie in different connected components of the cut
specified by $X^l$.

To compute the quality of an ultrametric $d$ with respect to an input set
of edge weights $\theta$, we measure the squared $L_2$ difference between 
the edge weights and the ultrametric distance $\|\theta - d\|_2^2$.
To write this compactly in terms of multicut indicator vectors, we construct a
set of weights for each edge and layer, denoted $\theta_e^l$ so that
$\sum_{l=0}^{m} \theta_e^l = \|\theta_e - \delta^m\|^2$. These weights are
given explicitly by the telescoping series:
\begin{equation}
\theta_e^0 = \|\theta_e\|^2 \quad \quad  
\theta_e^l =\|\theta_e - \delta^l\|^2 - \|\theta_e - \delta^{l-1}\|^2  \quad \forall l > 1 
\end{equation}
We use $\theta^l \in R^{|E|}$ to denote the vector containing $\theta^l_e$ for
all $e\in E$. 

For a fixed number of levels $L$ and fixed set of thresholds $\delta$, the
problem of finding the nearest ultrametric $d$ can then be written as an integer
linear program (ILP) over the edge cut indicators.
\begin{align}
\nonumber \min_{d} \sum_{e \in E} \| \theta_e - d_e \|^2 &= \min_{\mathcal{X} \in \bar{\Omega}_L} \sum_{e \in E} \left\| \theta_e - \sum^{L}_{l=0} \delta^l [\bar{X}_{e}^l > \bar{X}_{e}^{l+1}] \right\|^2 \\
\nonumber &= \min_{\mathcal{X} \in \bar{\Omega}_L} \sum_{e \in E} \sum^{L}_{l=0} \| \theta_e - \delta^l \|^2 (\bar{X}_{e}^l - \bar{X}_{e}^{l+1}) \\
\nonumber &= \min_{\mathcal{X} \in \bar{\Omega}_L} \sum_{e \in E} \left( \|\theta_e\|^2 \bar{X}_{e}^0 +\sum^L_{l=1} \left(\| \theta_e - \delta^l \|^2 - \| \theta_e - \delta^{l-1} \|^2\right) \bar{X}_{e}^l  +  \|\theta_e - \delta^L\|^2 \bar{X}_e^{L+1} \right) \\
\nonumber &= \min_{\mathcal{X} \in \bar{\Omega}_L}\sum_{l=0}^L  \sum_{e \in E} \theta^l_e \bar{X}^l_e\\
&= \min_{\mathcal{X} \in \bar{\Omega}_L}\sum_{l=0}^L  \theta^l \cdot \bar{X}^l 
\label{hiermulti}
\end{align}
This optimization corresponds to solving a collection of minimum-weight
multicut problems where the multicuts are constrained to be hierarchically 
consistent.

Computing minimum-weight multicuts (also known as correlation clustering) is NP
hard even in the case of planar graphs \cite{nphardplanar}.  A direct approach
to finding an approximate solution to Eq \ref{hiermulti} is to relax the
integrality constraints on $\bar{X}^l$ and instead optimize over the whole
polytope defined by the set of cycle inequalities.  We write $\cycpoly$ to
indicate the polytope of real valued indicator vectors $X$ that satisfying the
cycle inequalities
\begin{align}
\cycpoly = \left\{X \in [0,1]^{|E|} : \sum_{e \in c -\hat{e}} X_e \geq X_{\hat{e}}, \forall \! c \in C, \hat{e}\in c \right\}
\label{cycineq}
\end{align}
and use $\Omega_L$ to denote the corresponding relaxation of $\bar{\Omega}_L$
given by
\[
{\Omega}_L = \{ (X^1,X^2,\ldots X^L ) : X^l \in \cycpoly, X^l \geq X^{l+1} \; \forall l\}
\]
While the polytope $\cycpoly$ contains non-integral vertices (it is not the
convex hull of $\mcut$), the integral vertices of $\cycpoly$ do correspond
exactly to the set of valid multicuts \cite{deza1997geometry}.  

In practice, we found that applying a straightforward cutting-plane approach
that successively adds violated cycle inequalities to this relaxation of Eq
\ref{hiermulti} requires far too many constraints and is too slow to be useful.
Instead, we develop a column generation approach tailored for planar graphs
that allows for efficient and accurate approximate inference.

\section{The Cut Cone and Planar Multicuts}

Consider a partition of a planar graph into two disjoint sets of nodes.  We
denote the space of indicator vectors corresponding to such two-way cuts by
$\cut$.  A cut may yield more than two connected components but it can
not produce every possible multicut (e.g., it can not split a triangle of three
nodes into three separate components).  Let $Z \in \{ 0,1\}^{|E|\times
|\cut|}$ be an indicator matrix where each column specifies a valid
two-way cut with $Z_{ek}=1$ if and only if edge $e$ is cut in two-way cut $k$. 
The indicator vector of any multicut in a planar graph can be generated by a
suitable linear combination of of cuts (columns of $Z$) that isolate the
individual components from the rest of the graph where the weight of each such
cut is $\frac{1}{2}$.

Let $\gamma \in \mathbb{R}^{|\cut|}$ be a vector specifying a positive weighted
combination of cuts.  The set $\cutcone = \{Z\gamma : \gamma \geq 0\}$ is the
conic hull of $\cut$ or ``cut cone''. Since any multicut can be expressed as a
superposition of cuts, the cut cone is identical to the conic hull of $\mcut$.
This equivalence suggests an LP relaxation of the minimum-cost multicut given
by
\begin{equation}
\label{cutandhyperi}
\min_{\gamma \geq 0 }\theta \cdot Z\gamma \quad \quad s.t. \; \;  Z\gamma \leq 1
\end{equation}
where the vector $\theta \in \mathbb{R}^{|E|}$ specifies the edge weights.  For
the case of planar graphs, any solution to this LP relaxation satisfies the
cycle inequalities (see Appendix \ref{app1cyc} and \cite{deza1997geometry,
analpcc,Bar3}).

\begin{figure*}
\centering     
\subfigure[Linear combination of cut vectors]{\label{betaslackfig}
\includegraphics[width=66mm]{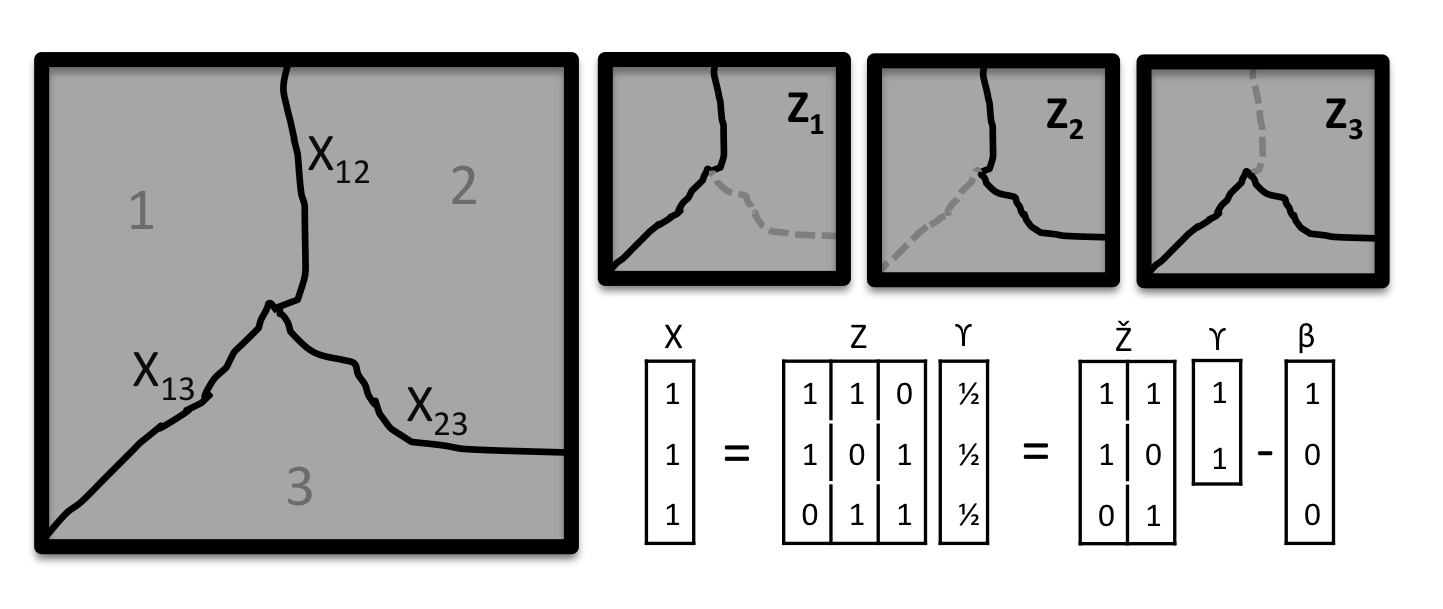}}
\subfigure[Hierarchical cuts]{\label{fixalpha1}
\includegraphics[width=66mm]{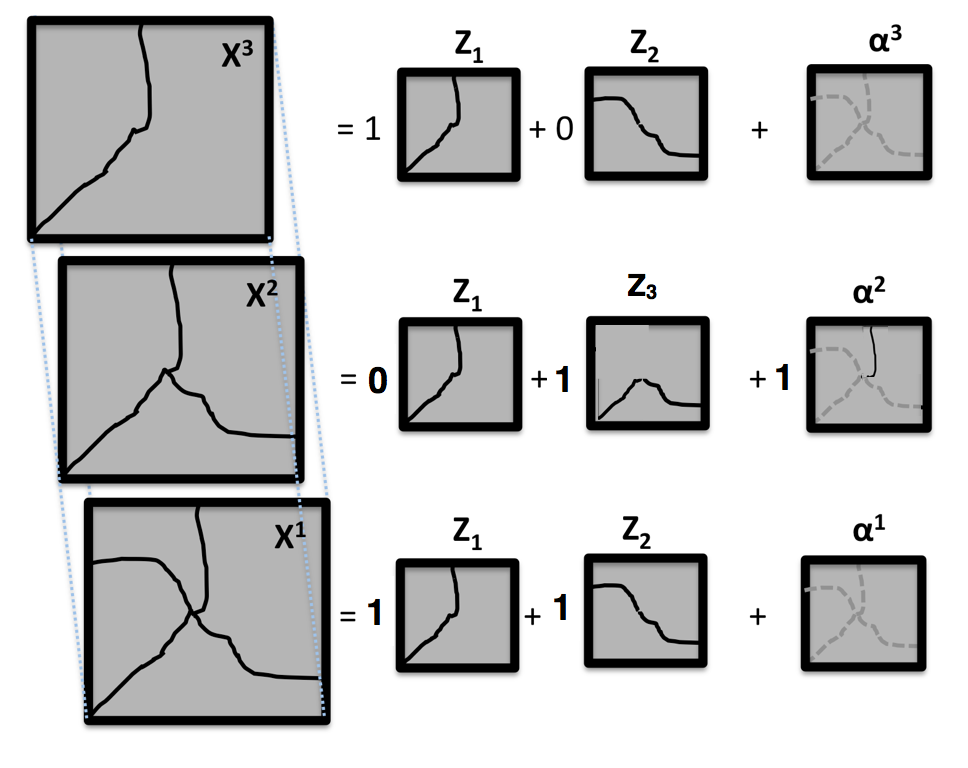}}
\caption{\textbf{(a)}  Any partitioning $X$ can be represented as a linear superposition
of cuts $Z$ where each cut isolates a connected component of the partition and
is assigned a weight $\gamma = \frac{1}{2}$ \cite{PlanarCC}.  By introducing an
auxiliary slack variables $\beta$, we are able to represent a larger set of
valid indicator vectors X using fewer columns of $Z$.  \textbf{(b)} By
introducing additional slack variables at each layer of the hierarchical
segmentation, we can efficiently represent many hierarchical segmentations
(here $\{X^1,X^2,X^3\}$) that are consistent from layer to layer while using
only a small number of cut indicators as columns of $Z$.
}
\end{figure*}

\noindent\textbf{Expanded Multicut Objective:}
Since the matrix $Z$ contains an exponential number of cuts, Eq.
\ref{cutandhyperi} is still intractable. Instead we consider an
approximation using a constraint set ${\hat Z}$ which is a subset of columns of
$Z$.  In previous work \cite{PlanarCC}, we showed that since the optimal
multicut may no longer lie in the span of the reduced cut matrix ${\hat Z}$, it
is useful to allow some values of ${\hat Z}\gamma$ exceed $1$ (see Figure
\ref{betaslackfig} for an example).  

We introduce a slack vector $\beta \geq 0$ that tracks the presence of any
``overcut'' edges and prevents them from contributing to the objective when the
corresponding edge weight is negative.  Let $\theta_e^-=\min(\theta_e,0)$
denote the non-positive component of $\theta_e$.  The expanded multi-cut
objective is given by:
\begin{equation}
\min_{\substack{\gamma \geq 0\\ \beta \geq 0}}\theta \cdot {\hat Z}\gamma-\theta^{-} \cdot  \beta \quad \quad 
s.t.\; \; {\hat Z}\gamma -\beta \leq 1 
\label{planareq}
\end{equation}
For any edge $e$ such that $\theta_e < 0$, any decrease in the
objective from overcutting by an amount $\beta_e$ it is exactly compensated for
in the objective by the term $-\theta^-_e \beta_e$.  

When ${\hat Z}$ contains all cuts (i.e., ${\hat Z}=Z$) then Eq \ref{cutandhyperi}
and Eq \ref{planareq} are equivalent \cite{PlanarCC}. Further, if $\gamma^\star$
is the minimizer of Eq \ref{planareq} when ${\hat Z}$ only contains a subset of
columns, then the edge indicator vector given by $X = \min(1,{\hat Z}\gamma^\star)$
still satisfies the cycle inequalities (see Appendix \ref{app1cyc} for details).

\section{Relaxing Ultrametric Rounding}

To relax the ultrametric rounding problem, we replace the multicut problem at
each layer $l$ using the expanded multicut objective described by Eq
\ref{planareq}.  We let $\gamma = \{\gamma^1,\gamma^2,\gamma^3\ldots \gamma^L
\}$ and $\beta = \{ \beta^1,\beta^2,\beta^3\ldots \beta^L \}$ denote the
collection of weights and slacks for the levels of the hierarchy and let
$\theta^{+l}_e=\max(0,\theta^{l}_e)$ and $\theta^{-l}_e=\min(0,\theta^{l}_e)$
denote the positive and negative components of $\theta^{l}$.  We write the
relaxed ultrametric rounding problem as:
\begin{align}
\label{imposversion}
\min_{\substack{ \gamma \geq 0 \\ \beta \geq 0 }}
   \sum^L_{l=1}& \left( \theta^l \cdot  Z \gamma^l- \theta^{-l}  \cdot \beta^l \right)\\
\nonumber s.t.\; \; &Z \gamma^{l+1} \leq Z\gamma^{l} \quad  \forall l < L\\
& Z\gamma^l - \beta^l\leq 1 \quad \forall l 
\end{align}
where we have dropped the $l=0$ term from Eq \ref{hiermulti} which is a
constant.

\noindent\textbf{Expanded Ultrametric Cut Cone Objective:}
As with Eq \ref{planareq}, it is computationally useful to introduce an
additional slack vector associated with each level $l$ and edge $e$ which we
denote as $\alpha = \{ \alpha^1,\alpha^2,\alpha^3 \ldots \alpha^{L-1}\}$.  
The introduction of $\alpha^l_e$ allows for cuts represented by $Z\gamma^l$ to
violate the hierarchical constraint $Z \gamma^{l}_e \geq Z\gamma^{l+1}_e$.
However we modify the objective so that violations to the original hierarchy
constraint are paid for in proportion to $\theta^{+l}_e$.  The introduction of
$\alpha$ allows us to find valid ultrametrics while using a smaller number of
columns of $Z$ to be used than would otherwise be required (illustrated in
Figure \ref{fixalpha1}).  We call this relaxed ultrametric rounding problem 
including the slack variable $\alpha$ the {\em expanded ultrametric rounding
objective}, written as:
\begin{align}
\label{primalultraz}
\min_{\substack{\gamma \geq 0 \\ \beta \geq 0 \\ \alpha \geq 0}}
  \sum^L_{l=1} & \theta^l \cdot Z \gamma^l + \sum^L_{l=1} -\theta^{-l} \cdot \beta^l + \sum^{L-1}_{l=1}  \theta^{+l} \cdot  \alpha^l &\\
s.t. \; \; \nonumber & Z\gamma^{l+1}+\alpha^{l+1} \leq Z \gamma^{l}+\alpha^l \quad \forall l <L\\
& Z\gamma^l-\beta^l \leq 1\quad \forall l
\end{align}
where by convetion we define $\alpha^{L} = 0$.

Given a solution $(\alpha,\beta,\gamma)$ we can recover a relaxed solution to the
ultrametric rounding problem (Eq. \ref{imposversion}) over $\Omega^L$ by setting
$X^l_e=\min(1,\max_{m \geq l}\;(Z\gamma^m)_e)$.  In  Appendix \ref{ProjectUltraha}, we
demonstrate that for any $(\alpha,\beta,\gamma)$ that obeys the constraints in
Eq \ref{primalultraz}, this thresholding operation yields a solution
$\mathcal{X}$ that lies in $\Omega^L$ and achieves the same or lower objective
value.

\section{The Dual Objective}
\label{duarderv}
We optimize the dual of the objective in Eq \ref{primalultraz} using an
an efficient column generation approach based on perfect matching.  A
detailed derivation is given in Appendix \ref{dualDerv}.  Briefly, We introduce two
sets of Lagrange multipliers $\omega =\{ \omega^1,\omega^2,\omega^3
\ldots \omega^{L-1} \}$ and $\lambda =\{ \lambda^1,\lambda^2,\lambda^3 \ldots
\lambda^L \}$ corresponding to the between and within layer constraints
respectively.  For notational convenience, let $\omega^0 = 0$.  The dual
objective can then be written as
\begin{align}
\label{dualprob}
\max_{\omega \geq 0, \lambda \geq 0 } & \sum_{l=1}^L -\lambda^l \cdot 1 \\
&\nonumber \theta^{-l} \leq -\lambda^l \quad \forall l \\
&\nonumber  -(\omega^{l-1}-\omega^{l}) \leq \theta^{+l} \quad \forall l\\
&\nonumber (\theta^l+\lambda^l +\omega^{l-1}-\omega^{l}) \cdot Z\geq 0 \quad \forall l 
\end{align}
The dual LP can be interpreted as finding a small modification of the original
edge weights $\theta^l$ so that every possible two-way cut of each resulting
graph at level $l$ has non-negative weight.  Observe that the introduction of
the two slack terms $\alpha$ and $\beta$ in the primal problem (Eq
\ref{primalultraz}) results in bounds on the Lagrange multipliers $\lambda$ and
$\omega$ in the dual problem in Eq \ref{dualprob}.  In practice these dual
constraints turn out to be essential for efficient optimization and constitute
the core contribution of this paper.

\section{Solving the Dual via Cutting Planes}

The chief complexity of the dual LP is contained in the constraints including $Z$
which encodes non-negativity of an exponential number of cuts of the graph
represented by the columns of $Z$.  To circumvent the difficulty of explicitly
enumerating the columns of $Z$, we employ a cutting plane method that
efficiently searches for additional violated constraints (columns of $Z$)
which are then successively added.  
 
Let $\hat{Z}$ denote the current working set of columns.  Our dual optimization
algorithm iterates over the following three steps: (1)  Solve the dual LP with
$\hat{Z}$, (2) find the most violated constraint  of the form
$(\theta^l+\lambda^l+\omega^{l-1}-\omega^{l}) \cdot Z \geq 0$ for layer $l$,  (3) Append a column to
the matrix $\hat{Z}$ for each such cut found.  We terminate when no violated
constraints exist or a computational budget has been exceeded. 

\subsection{Finding Violated Constraints}

Identifying columns to add to $\hat{Z}$ is carried out for each layer $l$
separately.  Finding the most violated constraint of the full problem
corresponds to computing the minimum-weight cut of a graph with edge weights
$\theta^l+\lambda^l+\omega^{l-1}-\omega^{l}$.  If this cut has non-negative
weight then all the constraints are satisfied, otherwise we add the corresponding
cut indicator vector as an additional column of $Z$.

To generate a new constraint for layer $l$ based on the current Lagrange
multipliers, we solve
\begin{align}
z^l= \arg\min_{z \in \cut} \sum_{e \in E}(\theta^l_e+\lambda^l_e+\omega^{l-1}_e-\omega^{l}_e) z_e
\end{align}
and subsequently add the new constraints from all layers to our LP, $\hat{Z} \leftarrow
[\hat{Z},\; z^1,\; z^2,\; \ldots\; z^L]$.  Unlike the  multicut problem,
finding a (two-way) cut in a planar graph can be solved exactly by a reduction
to minimum-weight perfect matching.  This is a classic result that, e.g.
provides an exact solution for the ground state of a 2D lattice Ising model
without a ferromagnetic field~\cite{fisher2,Bar1,Bar2,Bar3} in
$O(N^{\frac{3}{2}}\log N)$ time \cite{kolmblos}. 

\noindent{\bf Computing a lower bound:}
At a given iteration, prior to adding a newly generated set of constraints we
can compute the total residual constraint violation over all layers of hierarchy by $\Delta = \sum_l
(\theta^l+\lambda^l+\omega^{l-1}-\omega^{l}) \cdot z^l$.  In Appendix
\ref{genuinelow} we demonstrate that the value of the dual objective plus
$\frac{3}{2} \Delta$ is a lower-bound on the relaxed ultrametric rounding
problem in Eq \ref{primalultraz}.  Thus, as the costs of the minimum-weight
matchings approaches zero from below, the objective of the reduced problem over
${\hat Z}$ approaches an accurate lower-bound on optimization over
$\bar{\Omega}_L$
\subsection{Implementation Details}

{\bf Expanding generated cut constraints:}
When a given cut $z^l$ produces more than two connected components, we found it
useful to add a constraint corresponding to each component, following
the approach of \cite{PlanarCC}.  Let the number of connected components of
$z^l$ be denoted $M$.  For each of the  $M$ components then we add one column
to $Z$; one corresponding to the cut that isolates each connected component
from the rest. This allows more flexibility in representing the final optimum
multicut as superpositions of these components.  In addition, we also found it
useful in practice to maintain a separate set of constraints $\hat{Z}^l$ for
each layer $l$.  Maintaining independent constraints $\hat{Z}^1, \hat{Z}^2,
\ldots, \hat{Z}^L$ can result in a smaller overall LP.

{\bf Speeding convergence of $\omega$:}
We found that adding an explicit penalty term to the objective
that encourages small values of $\omega$ speeds up convergence dramatically
with no loss in solution quality.  This penalty is scaled by a parameter
$\epsilon=10^{-4}$ which is chosen to be extremely small  in magnitude  relative to the values
of $\theta$ so that it only has an influence when other no other ``forces'' are
acting on a given term in  $\omega$.  With this refinement, the LP solved at
each iteration of the cutting plane algorithm is given as follows.  
\begin{align}
\label{findual}
\max_{\omega \geq 0, \lambda \geq 0} & \sum_{l =1}^{L} -\lambda^l   1 - \epsilon \|\omega\|_1 \\
s.t. \; \;
&\nonumber \theta^{-l} \leq -\lambda^l \quad \forall l \\
&\nonumber  -(\omega^{l-1}-\omega^{l}) \leq \theta^{+l} \quad \forall l\\
&\nonumber (\theta^l+\lambda^l +\omega^{l-1}-\omega^{l})Z\geq 0 \quad \forall l 
\end{align}

\begin{algorithm}
\caption{Dual Ultrametric Rounding via Cutting Planes}
\begin{algorithmic} 
  \STATE $\hat{Z}^l\leftarrow \{ \} \quad \forall l,  \quad \mbox{residual} \leftarrow -\infty$
  \WHILE{$\mbox{residual}<0$}
    \STATE $ \{ \omega \}, \{ \lambda \} \leftarrow$ Solve Eq \ref{findual} given $\hat{Z}$ 
    \STATE $\mbox{residual}=0$
    \FOR{$l=1:L $}
      \STATE $z^l \leftarrow \arg\min_{z \in \cut} (\theta^l+\lambda^l+\omega^{l-1}-\omega^{l}) \cdot z$
      \STATE $\mbox{residual} \leftarrow \mbox{residual} + \frac{3}{2} (\theta^l+\lambda^l+\omega^{l-1}-\omega^{l}) \cdot z^l$
      \STATE $\{z(1),z(2),\ldots,z(M)\} \leftarrow \mbox{isocuts}(z^l)$
      \STATE $\hat{Z}^l \leftarrow \hat{Z}^l \cup \{z(1),z(2),\ldots,z(M)\}$
    \ENDFOR
  \ENDWHILE
\end{algorithmic}
  \label{dualsolve}
\end{algorithm}
\subsection{Primal Decoding}
Algorithm \ref{dualsolve} gives a summary of the dual solver which at termination
produces a lower-bound as well as a set of cuts described by the constraint
matrices $\hat{Z}^l$. The subroutine {\em $\mbox{isocuts}(z^l)$} computes the
set of cuts that isolate each connected component of $z^l$

To generate a hierarchical clustering, we solve the primal, Eq
\ref{primalultraz}, using this reduced set $\hat{Z}$ in order to
recover a fractional solution $X^l_e=\min(1,\max_{m \geq l}(\hat{Z}^m\gamma^m
)_e)$.  We use an LP solver (IBM CPLEX) which provides this primal solution
``for free'' when solving the dual in Algorithm 1.

We round this fractional solution to a discrete hierarchical clustering using a
simple thresholding strategy.  We threshold the fractional $X$ as follows:
$\bar{X}^l_e \leftarrow [X^l_e > t]$.  We then repair any cut edges that lie
inside a connected component by setting them to zero to assure that $\bar{X}^l
\in \mcut$.  In our implementation we test a few discrete thresholds $t \in
\{0,0.2,0.4,0.6,0.8\}$ and take that threshold that yields $\bar{X}$ with the
lowest cost.  After each pass through the loop of Alg. \ref{dualsolve}  we
compute these upper-bounds and retain the optimum solution observed thus far.  

\section{Experiments}
We applied our algorithm on segmentation problems based on images from the
Berkeley Segmentation Data set (BSDS) \cite{bsdspaper}. To construct our input
graph we use superpixels generated by performing an oriented watershed
transform on the output of the global probability of boundary (gPb) edge
detector \cite{gpbpaper}.  The vertices of the graph are superpixels and edges
connect superpixels that are neighbors in the image, yielding a planar graph.
 
We construct base distance costs $\theta$ by using the log-odds ratio of
the local estimate of boundary contrast given by averaging $gPb$ classifier
output over the boundary between neighboring superpixels to yield a value
$gPb_e$. We truncated extreme values to enforce that $gPb_e \in
[\epsilon,1-\epsilon]$ with $\epsilon = 0.001$.  We set $\theta_e =
\log\left(\frac{1-gPb_e}{gPb_e}\right) +
\log\left(\frac{1-\epsilon}{\epsilon}\right)$
The additive offset assures that $\theta_e \geq 0$.
In our experiments we use a fixed set of eleven distance threshold levels
$\{\delta_l\}$ that uniformly spanned the useful range of threshold values
$[9.6,12.6]$.
We weighted edges proportionally to the length of the corresponding boundary in
the image.  We performed dual cutting plane iterations until convergence or
2000 seconds had passed.  Lower-bounds for the BSDS segmentations were on the
order of $-10^3$ or $-10^4$.  We terminate when the total residual is greater
than $-2\times10^{-4}$.  All codes were written in MATLAB using the Blossom V
implementation of minimum-weight perfect matching \cite{kolmblos} and the IBM
ILOG CPLEX LP solver with default options.

\subsection{Qualitative and Quantitative Results on Images}
\label{startExper}
\begin{figure*}
\centering     
\subfigure{\includegraphics[width=30mm]{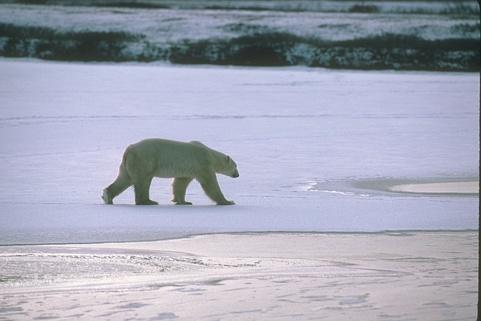}}
\subfigure{\includegraphics[width=30mm]{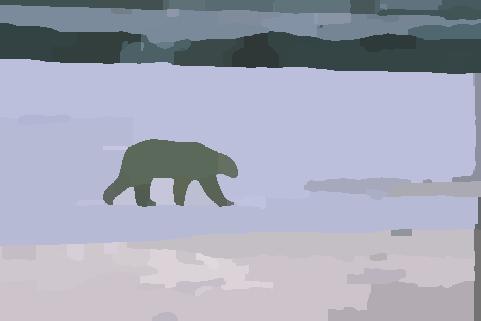}}
\subfigure{\includegraphics[width=30mm]{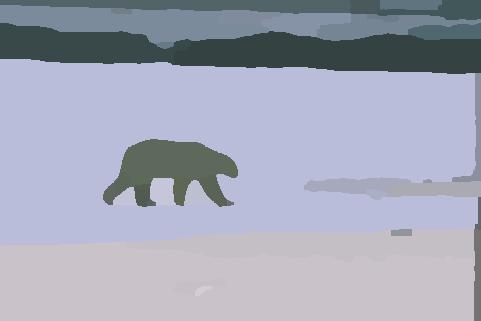}}
\subfigure{\includegraphics[width=30mm]{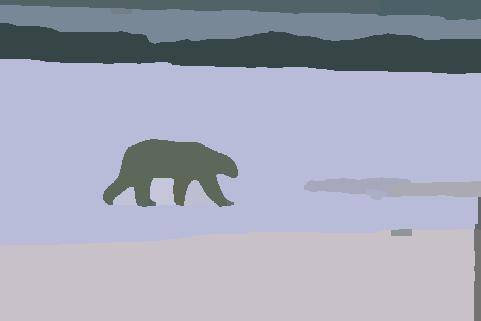}}
\subfigure{\includegraphics[width=30mm]{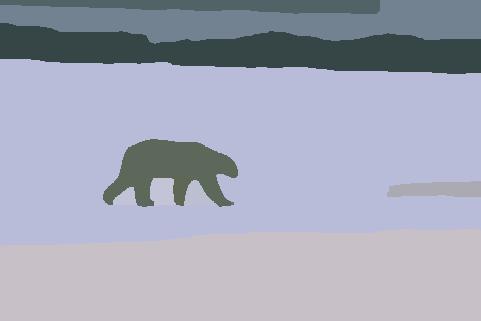}}
\subfigure{\includegraphics[width=30mm]{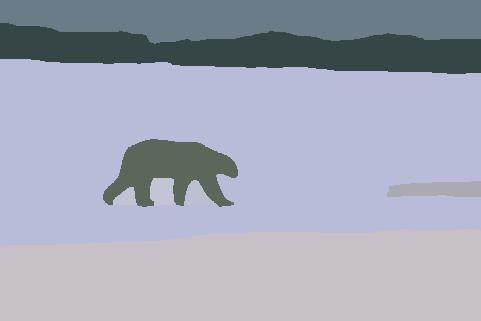}}
\subfigure{\includegraphics[width=30mm]{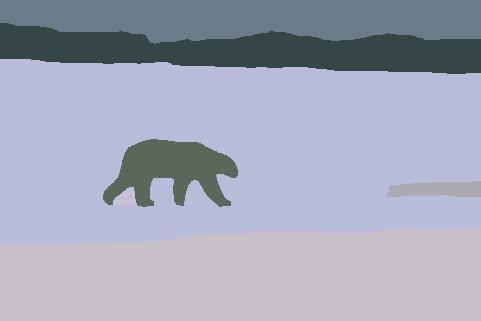}}
\subfigure{\includegraphics[width=30mm]{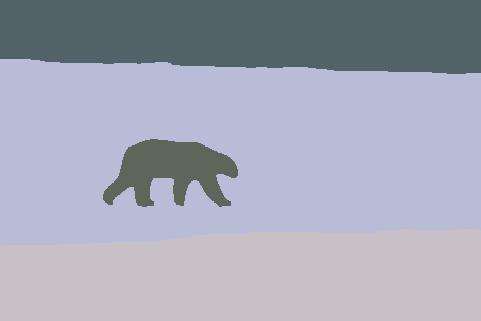}}
\subfigure{\includegraphics[width=30mm]{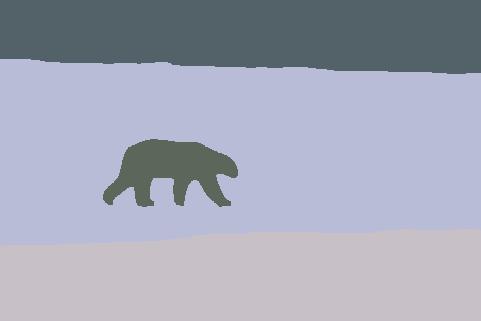}}
\subfigure{\includegraphics[width=30mm]{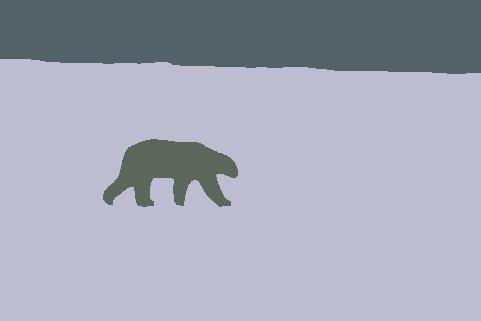}}
\subfigure{\includegraphics[width=30mm]{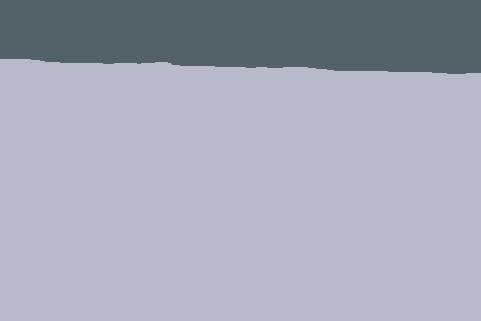}}
\subfigure{\includegraphics[width=30mm]{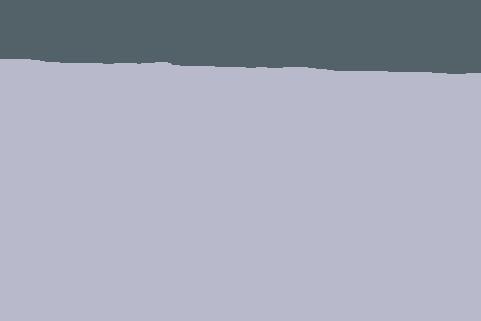}}
\caption{Top left to bottom right:  A hierarchical image segmentation for a
BSDS test set image showing eleven layers listed from fine to coarse.  The
original image is in the top left.  }
\label{qual1}
\end{figure*}

\begin{figure*}
\centering     
\subfigure{\includegraphics[width=30mm]{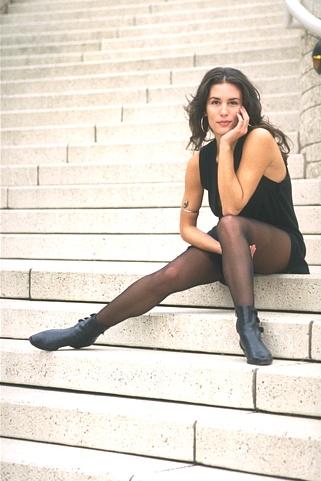}}
\subfigure{\includegraphics[width=30mm]{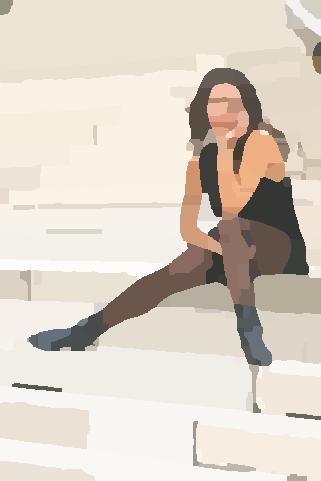}}
\subfigure{\includegraphics[width=30mm]{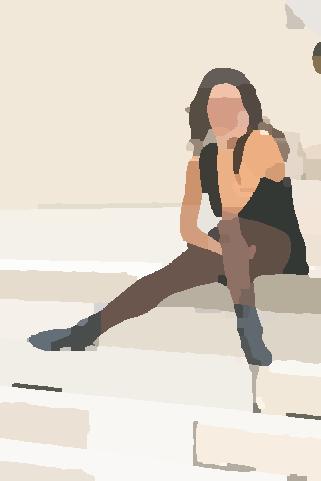}}
\subfigure{\includegraphics[width=30mm]{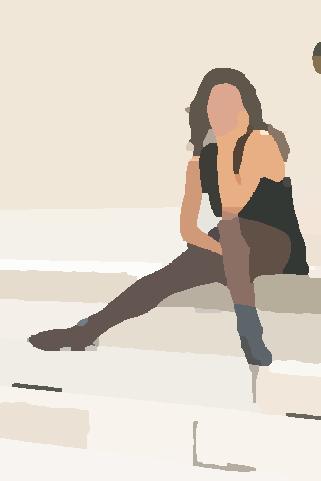}}
\subfigure{\includegraphics[width=30mm]{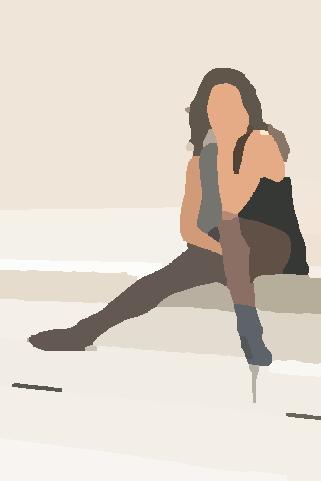}}
\subfigure{\includegraphics[width=30mm]{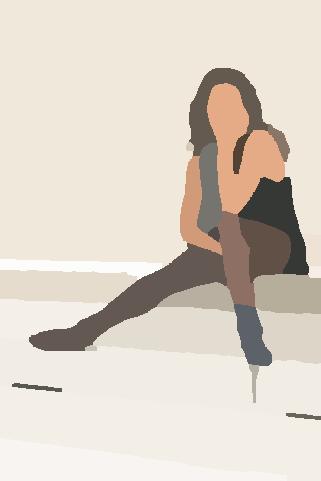}}
\subfigure{\includegraphics[width=30mm]{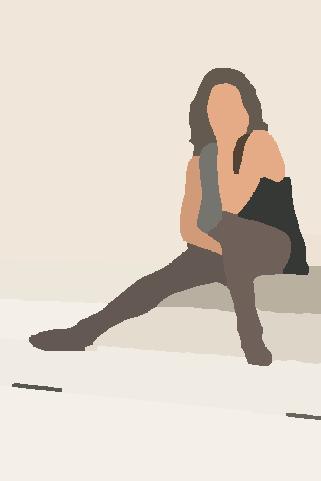}}
\subfigure{\includegraphics[width=30mm]{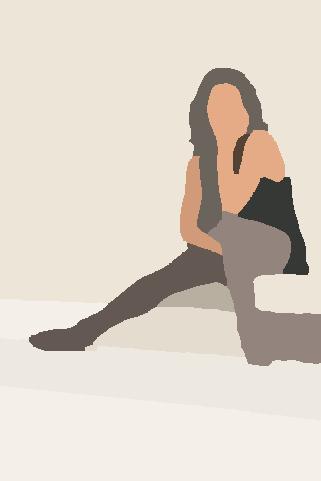}}
\subfigure{\includegraphics[width=30mm]{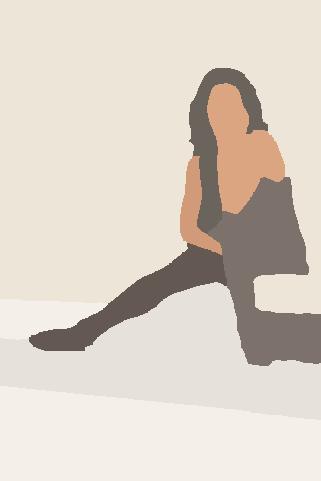}}
\subfigure{\includegraphics[width=30mm]{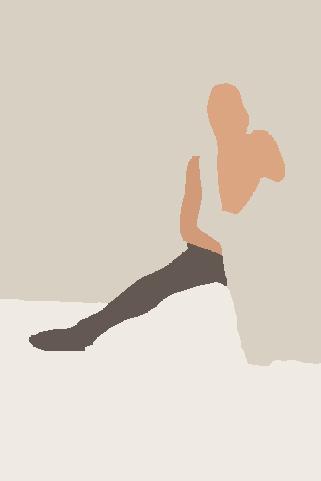}}
\subfigure{\includegraphics[width=30mm]{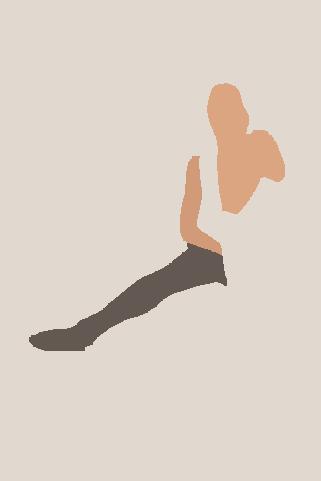}}
\subfigure{\includegraphics[width=30mm]{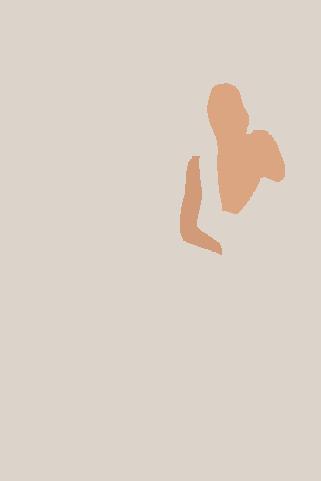}}
\caption{Top left to bottom right:  A hierarchical image segmentation for a
BSDS test set image showing eleven layers listed from fine to coarse.  The
original image is in the top left.  }
\label{qual2}
\end{figure*}

\begin{figure}
	\centering
  \includegraphics[width=0.6\linewidth,clip,trim=5mm 60mm 5mm 60mm]{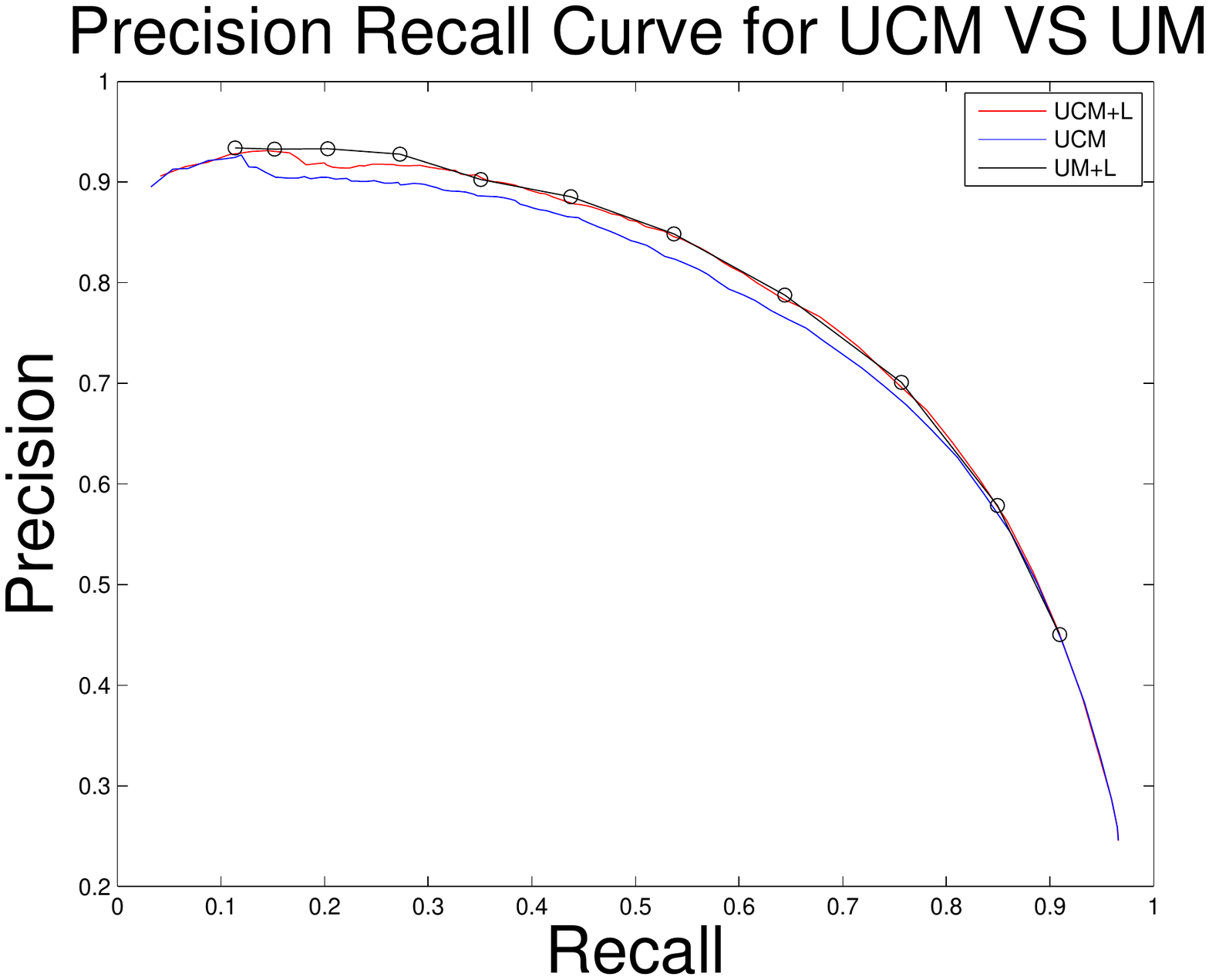}
\caption{
We show the comparison of our ultrametric rounding algorithm (UM) with the
baseline ultrametric contour maps algorithm  (UCM) with and without length
weighting \cite{UCM}.  We display precision recall plots on the Berkeley
Segmentation Data Set (BSDS),  Observe that UM performs nearly identically to
the state of the art UCM algorithm with regards to precision recall.  However
we do observe small but significant  improvements in high precision range of
the curve.  We note the points plotted on the precision recall curve for UM
with black dots.  Use of length weighted costs are indicated by $+L$.
}
\label{ultraper}
\end{figure}

Figs \ref{qual1}, \ref{qual2} show qualitative results for two images  from the
BSDS test data set.  We display segmentations at eleven thresholds and color
connected components of the segmentation at each layer with the average pixel
color over that component.  In Fig \ref{ultraper} we show the comparison of our
ultrametric rounding algorithm (Alg 1,denoted UM) with the baseline ultrametric
contour maps algorithm (UCM) with and without length weighting \cite{UCM}.
UCM performs agglomerative clustering algorithm, successively merging segments
with small boundary strengths to produce a hierarchical segmentation.  We
display a precision recall plot on the Berkeley Segmentation Data Set test set. 

In terms of segmentation accuracy, UM rounding performs nearly identically to
the state of the art UCM algorithm with regards to precision recall which is
the standard measure employed in the literature.  However we show some
improvements in high precision range of the curve which corresponds to the
coarse segmentations.  It is worth noting that the BSDS benchmark does not
provide strong penalties for small leaks between two segments when the total
number of boundary pixels involved is small.  Our algorithm may find strong
application in domains where the local boundary signal is noisier (e.g.,
biological imaging) or when under-segmentation is more heavily penalized.

\subsection{Objective Cost and Timing Experiments}

\begin{figure*}
\centering     
\subfigure[]{\label{quant1a:a}\includegraphics[width=65mm,clip,trim=5mm 60mm 5mm 60mm]{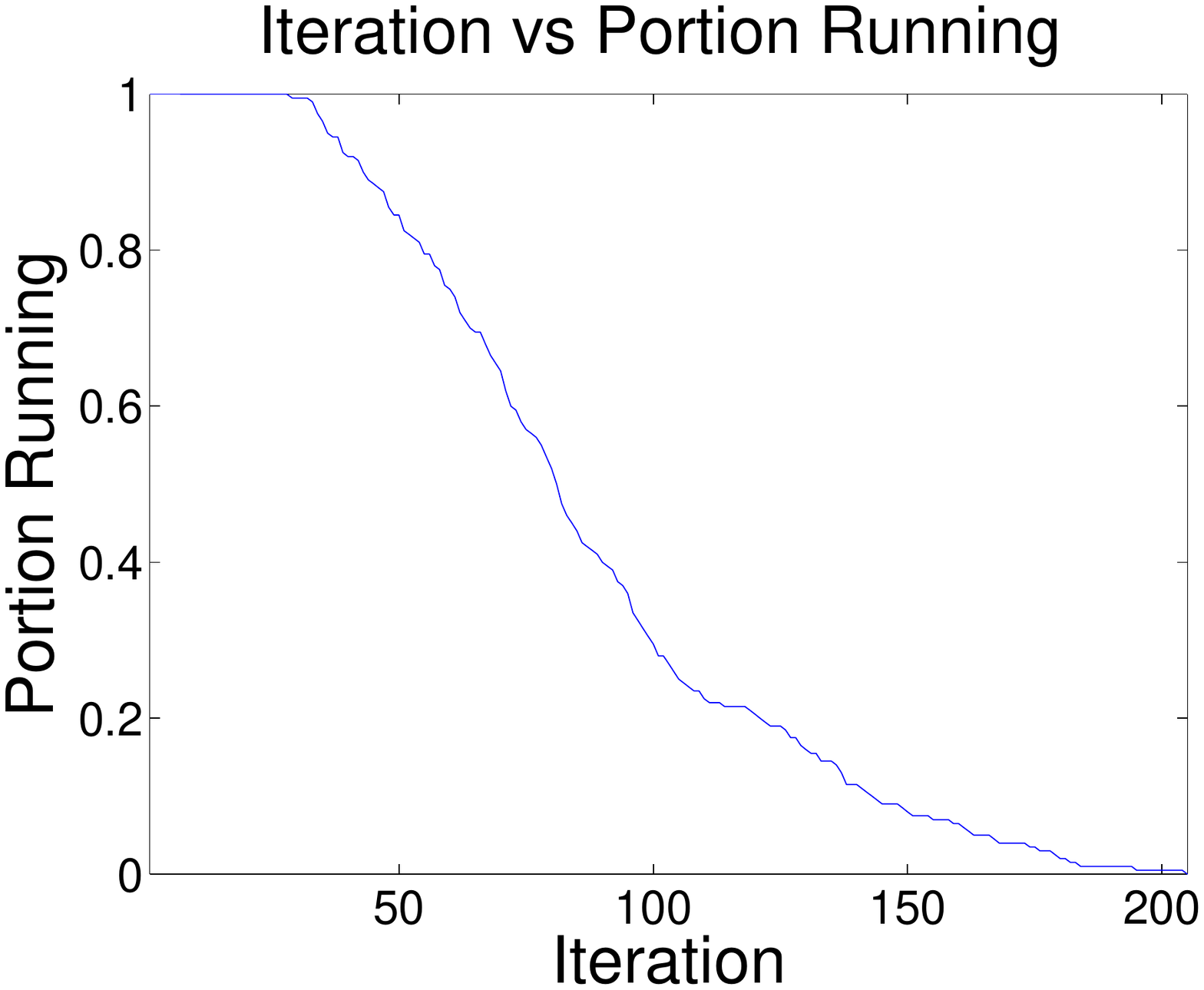}}
\subfigure[]{\label{quant1a:b}\includegraphics[width=65mm,clip,trim=5mm 60mm 5mm 60mm]{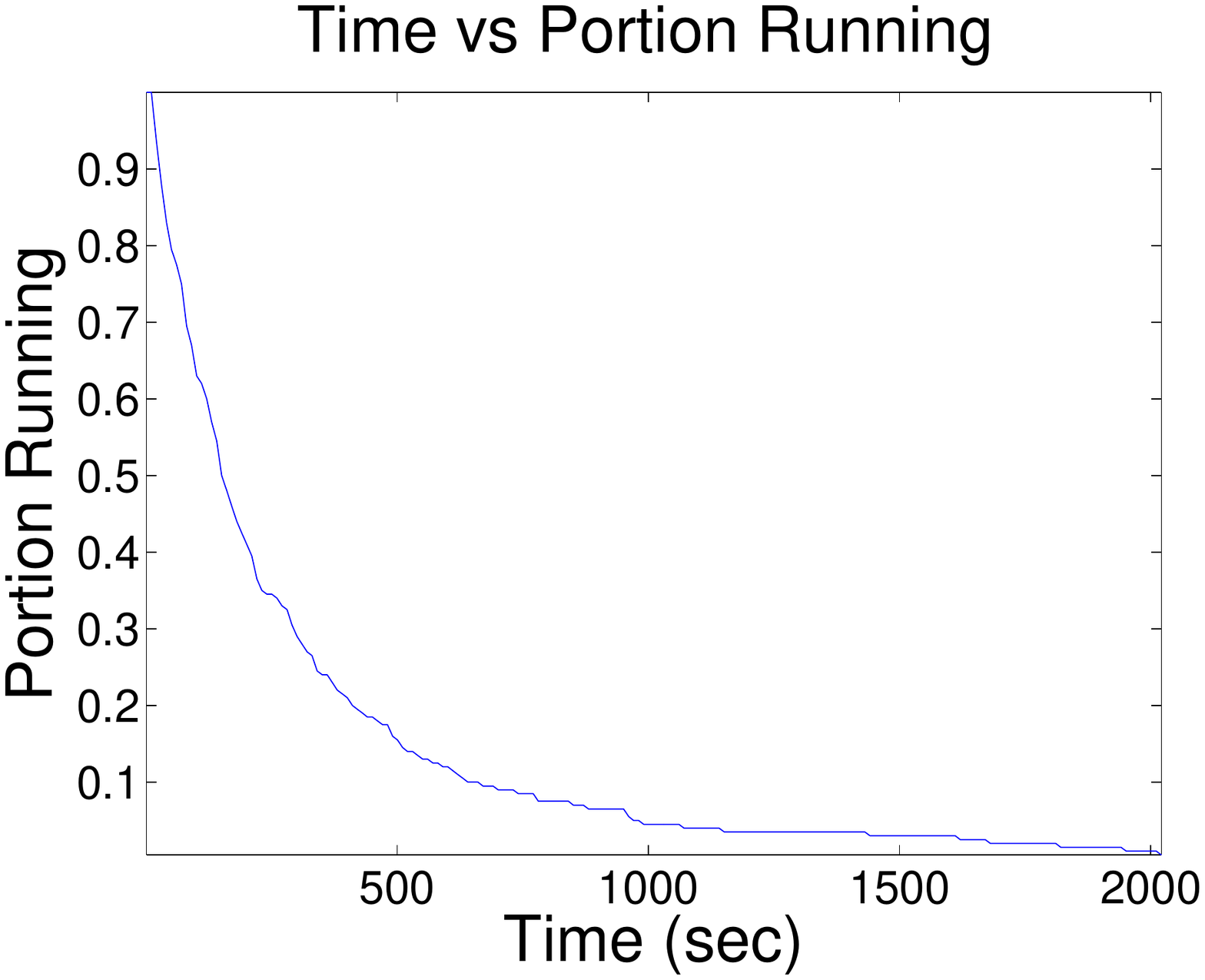}}
\subfigure[]{\label{quant1a:c}\includegraphics[width=65mm,clip,trim=5mm 60mm 5mm 60mm]{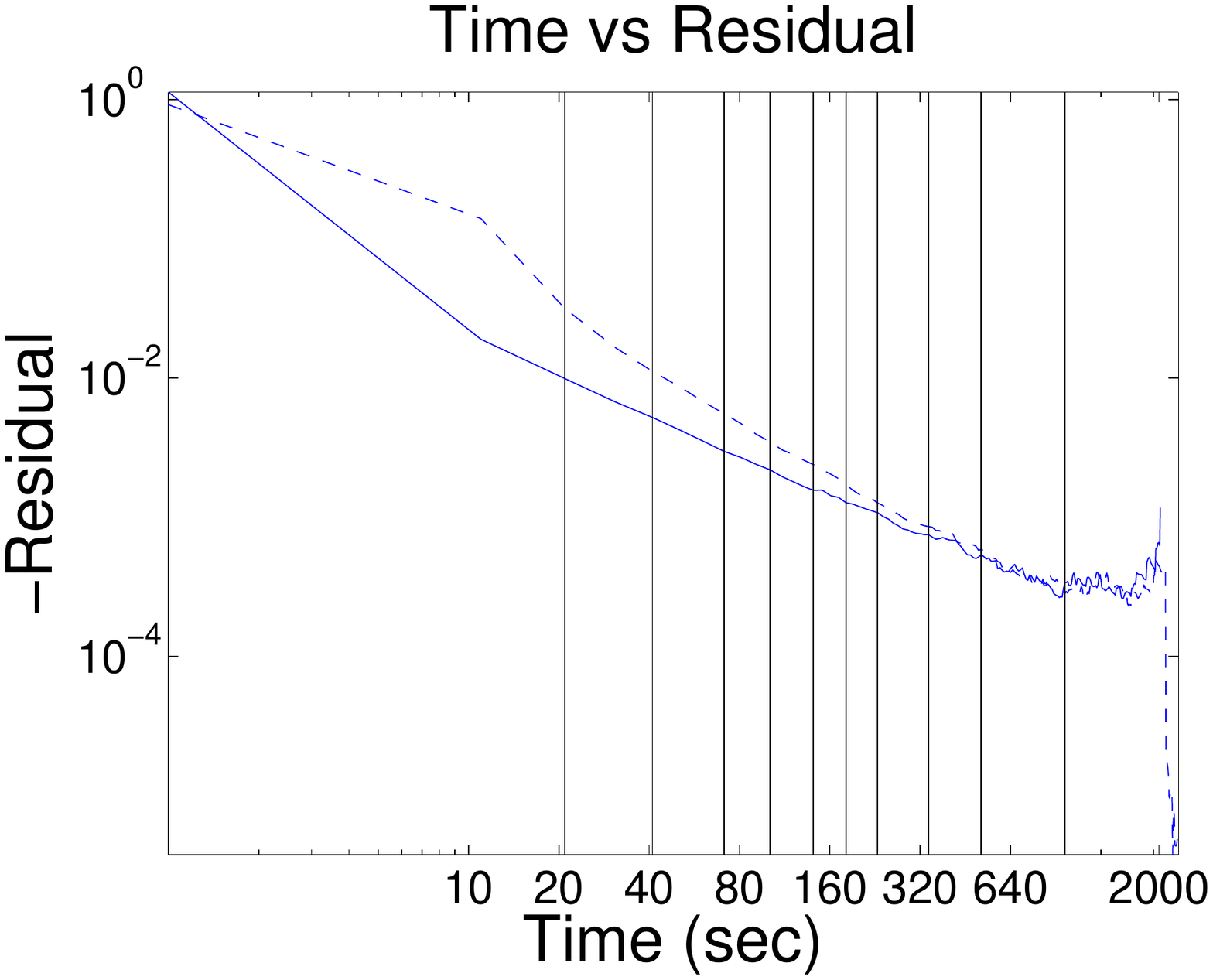}}
\subfigure[]{\label{quant1a:d}\includegraphics[width=65mm,clip,trim=5mm 60mm 5mm 60mm]{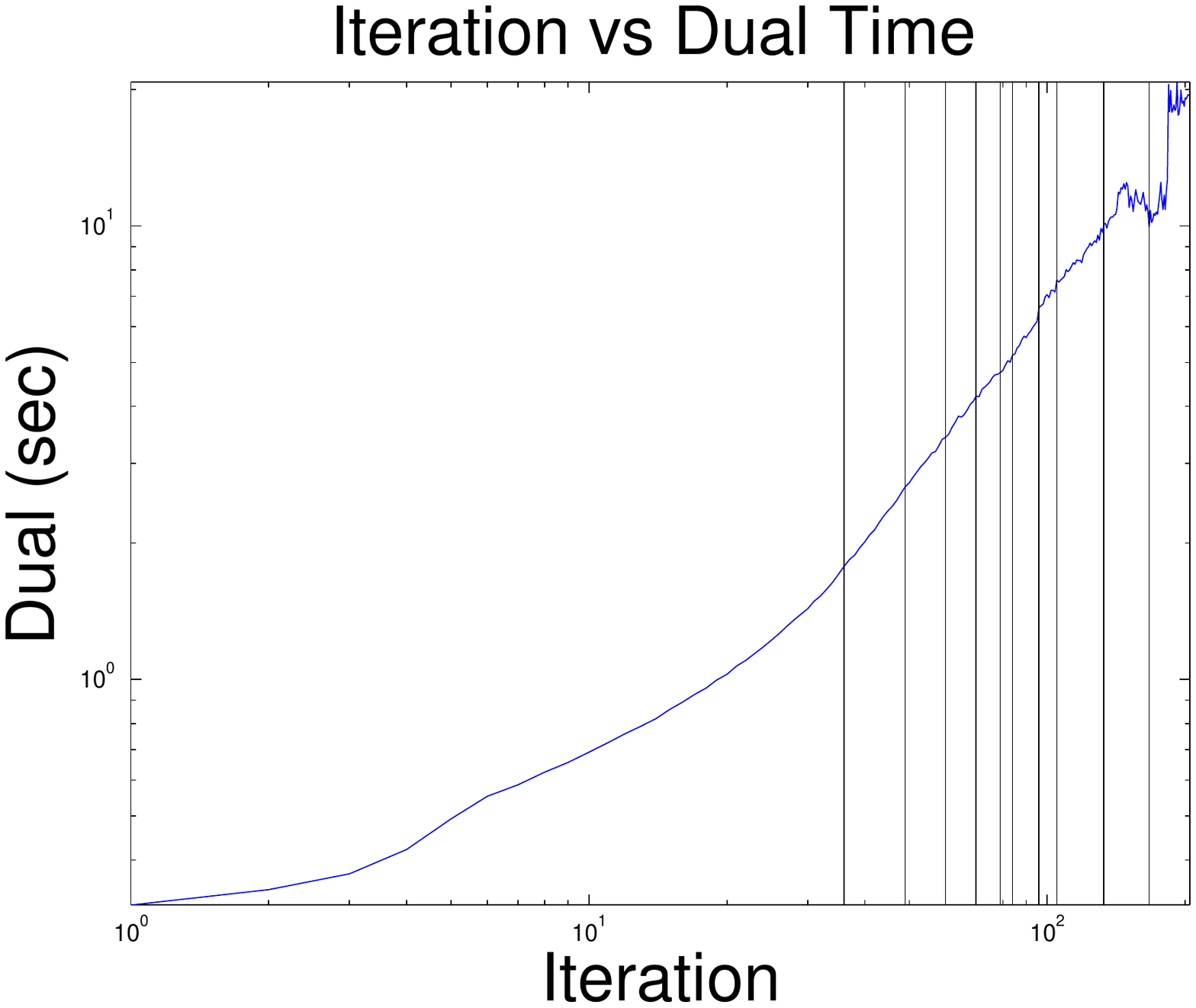}}
\caption{\textbf{(a)}  We display the portion of the problems that have not
terminated as a function of cutting-plane iteration.  We observe that dual
optimization always requires the solution of at least a few LP's for most
problems to converge. \textbf{(b)} We display the portion of the problems
that have not terminated as a function of time.  We observe that dual
optimization terminates rapidly for most problem instances.
\textbf{(c)}  We plot the value of the average residual constraint violation
as a function of time
averaged over images that have yet to terminate.  Instances that terminated
before 2000 seconds passed have residuals on the order of $10^{-6}$ or less.
We plot the best observed value in solid blue and the current value with dotted
blue. We normalize the residual for a given instance by dividing by the
magnitude of the tightest lower bound for that instance.  We indicate the
portion of instances that have yet to terminate using black bars.  The bars are
associated with the percent of instances incomplete with the bars from left to
right being [95,85,75,65,.....5]. 
Observe that the value of the residual decays rapidly. \textbf{(d)} We
plot the average amount of time per cutting-plane iteration.
This includes solving one LP and
finding the most violated constraint and extracting cuts for each layer.  We use
black bars as in \textbf{(c)} to indicate the percent of problems instances that
have not terminated after a given time point.}
\label{quant1a}
\end{figure*}

\begin{figure*}
\centering     
\subfigure[]{\label{quant1b:a}\includegraphics[width=65mm,clip,trim=5mm 60mm 5mm 60mm]{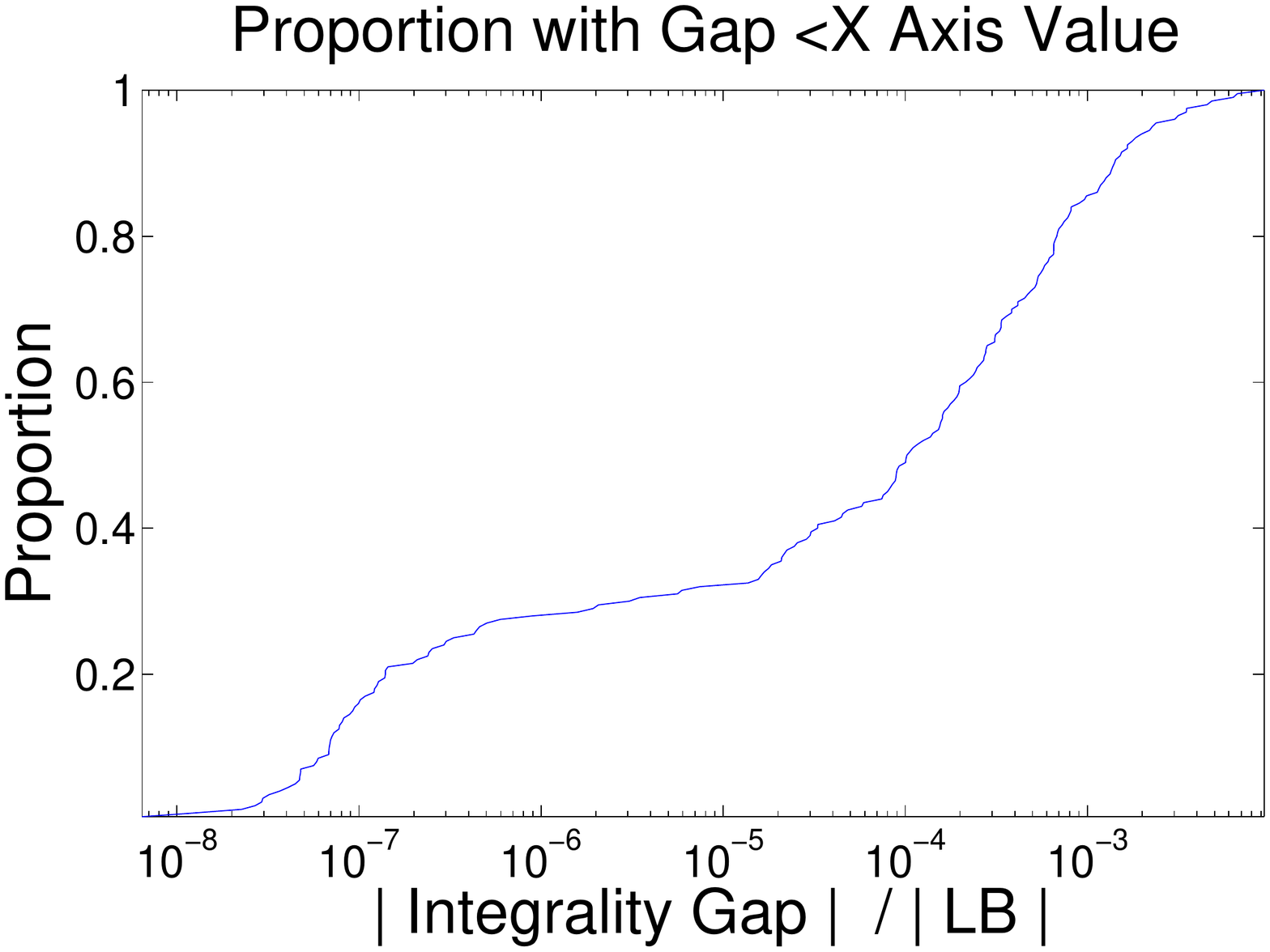}}
\subfigure[]{\label{quant1b:b}\includegraphics[width=65mm,clip,trim=5mm 60mm 5mm 60mm]{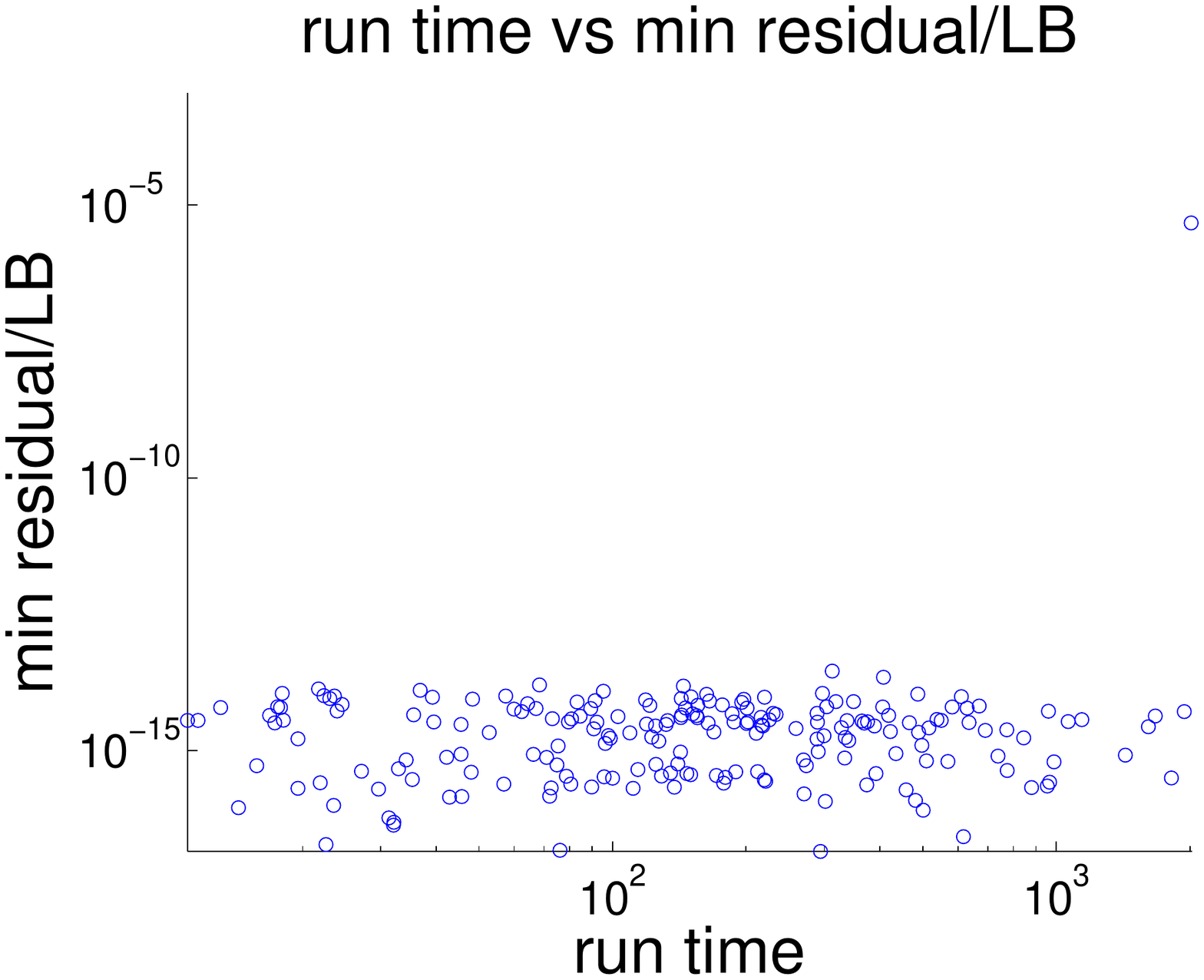}}
\subfigure[]{\label{quant1b:c}\includegraphics[width=65mm,clip,trim=5mm 60mm 5mm 60mm]{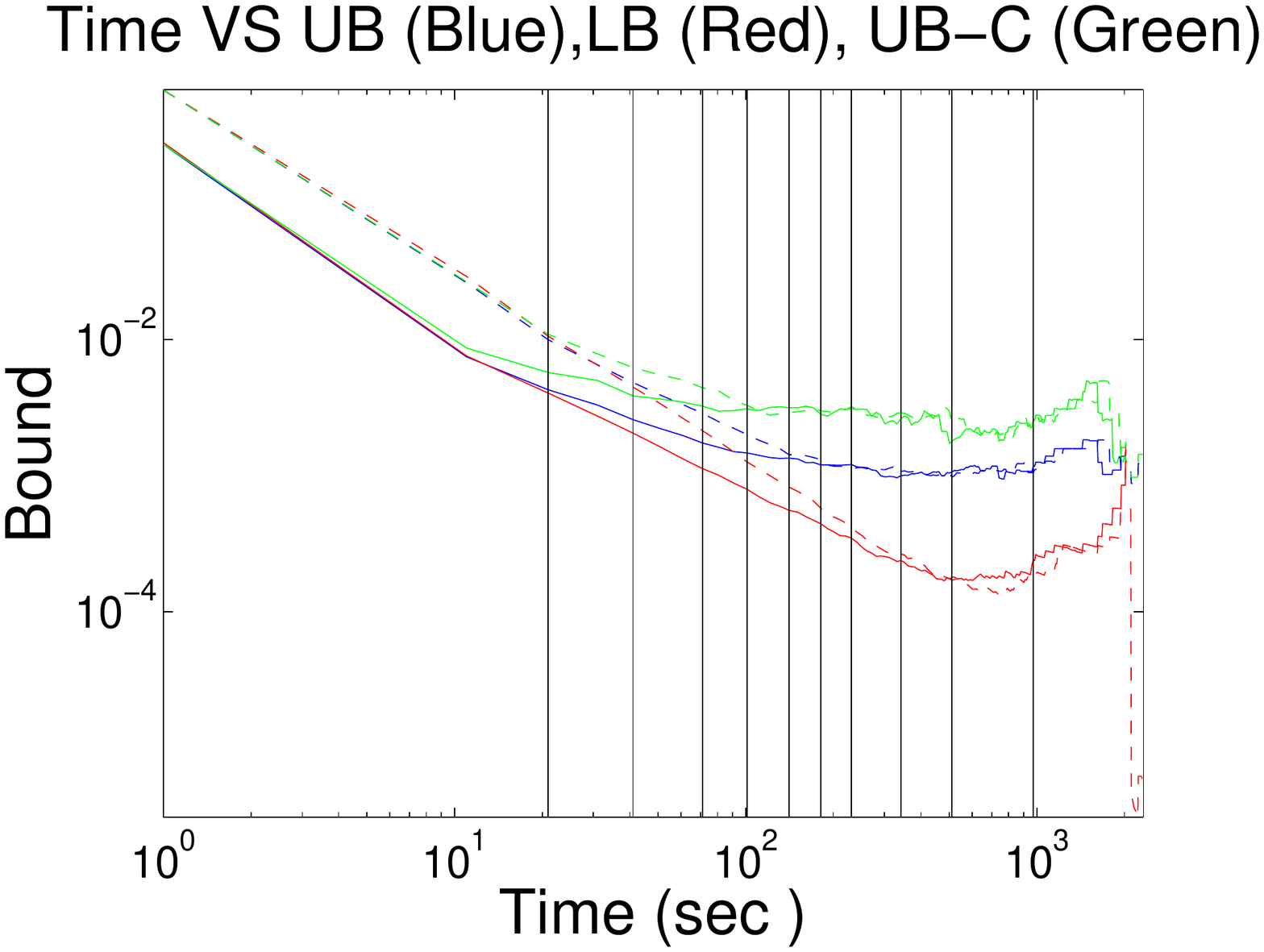}}
\subfigure[]{\label{quant1b:d}\includegraphics[width=65mm,clip,trim=5mm 60mm 5mm 60mm]{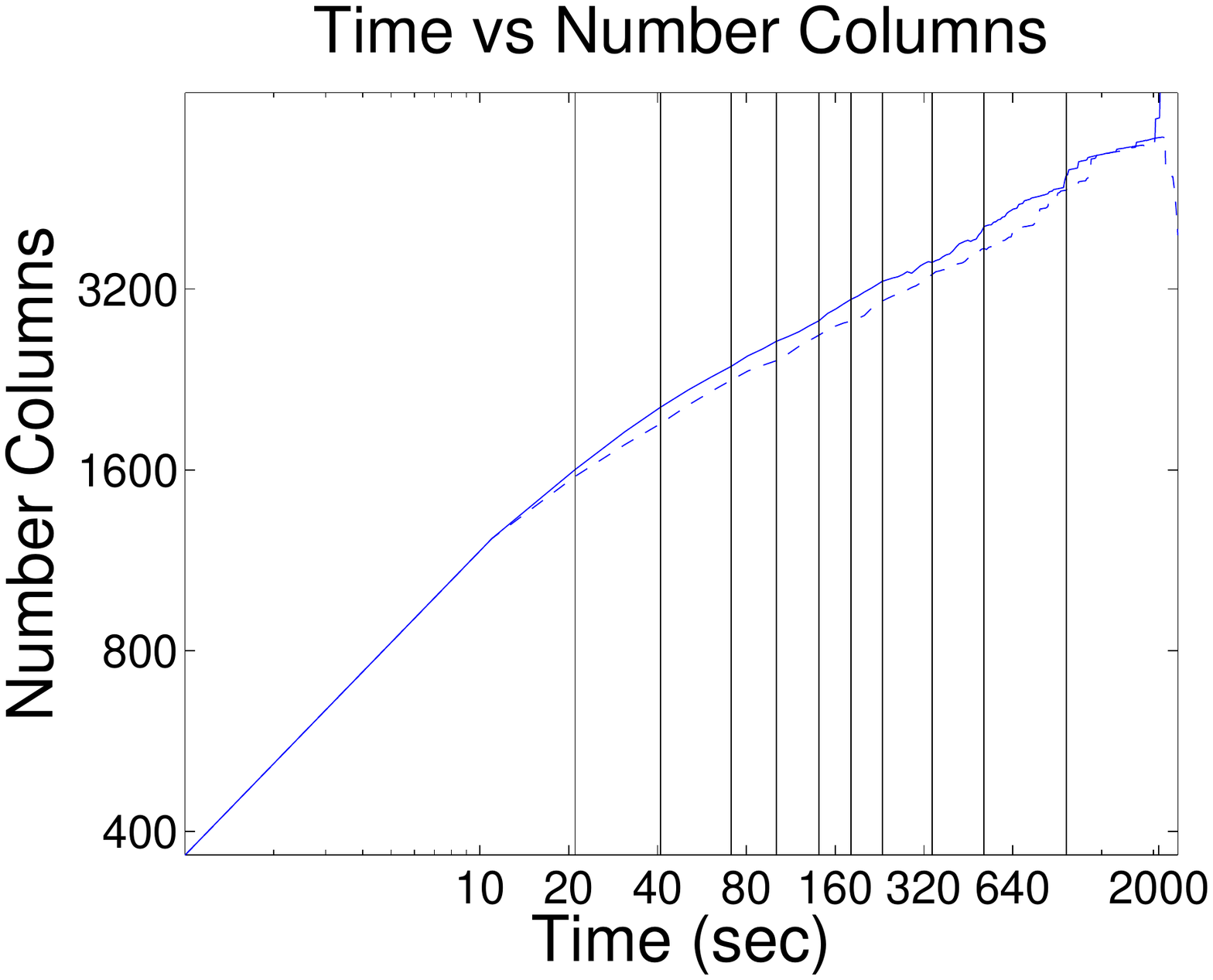}}
\caption{\textbf{(a)} We show a plot over image problem instances
that describes the gap between the maximum lower bound computed for that image
and the final rounded integral solution.  To normalize the energy gap, we scale
by the value of the maximum lower bound identified for that problem instance.
We observe that the rounded integer solutions are near exact or exact on all
images. 
\textbf{(b)} Scatter plot of the run time in (sec) versus the minimum
magnitude residual (residual is always non-positive).  We normalize this by
dividing by the maximum lower bound over the coarse optimization (denoted LB)
of the problem instance. Residual was negligible for all except 1 of 200
problem instances which did not terminate within 2000 seconds.
 \textbf{(c)}:  We show the value of the integer solution and  lower bound as
 a function of time averaged over problem instances.  We normalize by computing
 the absolute value of the gap between each bound and the magnitude of the
 maximum lower bound discovered.  We plot the value of the upper/lower bounds
 in blue/red.  We plot in green the value of the integer solution but include
 time for rounding the solution after each iteration.   We use dotted/solid
 lines to indicate the current/best value observed thus far.  We indicate the
 percentage of instances that have yet to terminate using black bars marking
 $[95,85,75,65,.....5]$ percent.  \textbf{(d)} We show the number of constraints
 (columns of $\hat{Z}^l$ summed over layers and averaged over problem instances) as a function of 
 running time.  We use black bars as in \textbf{c} to indicate the proportion
 of the problems instances that have not converged at a given time point.  
}
\label{quant1b}
\end{figure*}

In Fig \ref{quant1a},\ref{quant1b},we display plots demonstrating the
performance of the optimization routine according to eight different measures.
The most interesting is the quality of the integer solution.  We found the
upper-bound given by the cost of the decoded integer solution and the
lower-bound estimated by the dual LP are very close. The magnitude of the integrality gap  is
typically less than 0.1\% of the magnitude of the lower-bound and never more than $1$\%.
Convergence of the dual is achieved quite rapidly; most instances require less
than 100 iterations to converge with roughly linear growth in the size of the
LP at each iteration as cutting planes are added.

\subsection{Cost Comparison with Ultrametric Contour Maps}

We also compared the ultrametric rounding cost of solutions generated by our
approach with costs associated with hierarchical clusterings produced by the
Ultrametric Contour Map (UCM) length-weighted clusterings.  This test is
perhaps unfair as UCM was not necessarily designed to minimize the ultrametric
rounding cost but provides a baseline for understanding the rounding objective.

UCM provides an ultrametric solution denoted $U \in R^{|E|}$ where U is
indexed by $e$ and scaled to lie in the range $[0,1]$ with smaller values
indicating lower likelihood of a boundary.  For each level $l$, we select a
threshold $q \in [0,1]$ which is used to threshold the UCM ultrametric $U$.
We choose a value for $q$ which minimizes the ultrametric rounding error,
formally written as:
\begin{align}
\min_{q^l}\sum_{e \in E} \theta^l_e [U_e>q^l]
\end{align}
Thus the total cost for a given image is:
\begin{align}
\sum_{l=1}^{L} \min_{q^l}\sum_{e \in E} \theta^l_e [U_e>q^l]
\end{align}
Observe that  $\theta^l_e<\theta^{l+1}_e$ and thus we are guaranteed $q^l_e\leq
q^{l+1}_e$.

\begin{figure}
\centering
\includegraphics[width=0.6\linewidth,clip,trim=0mm 0mm 0mm 0mm]{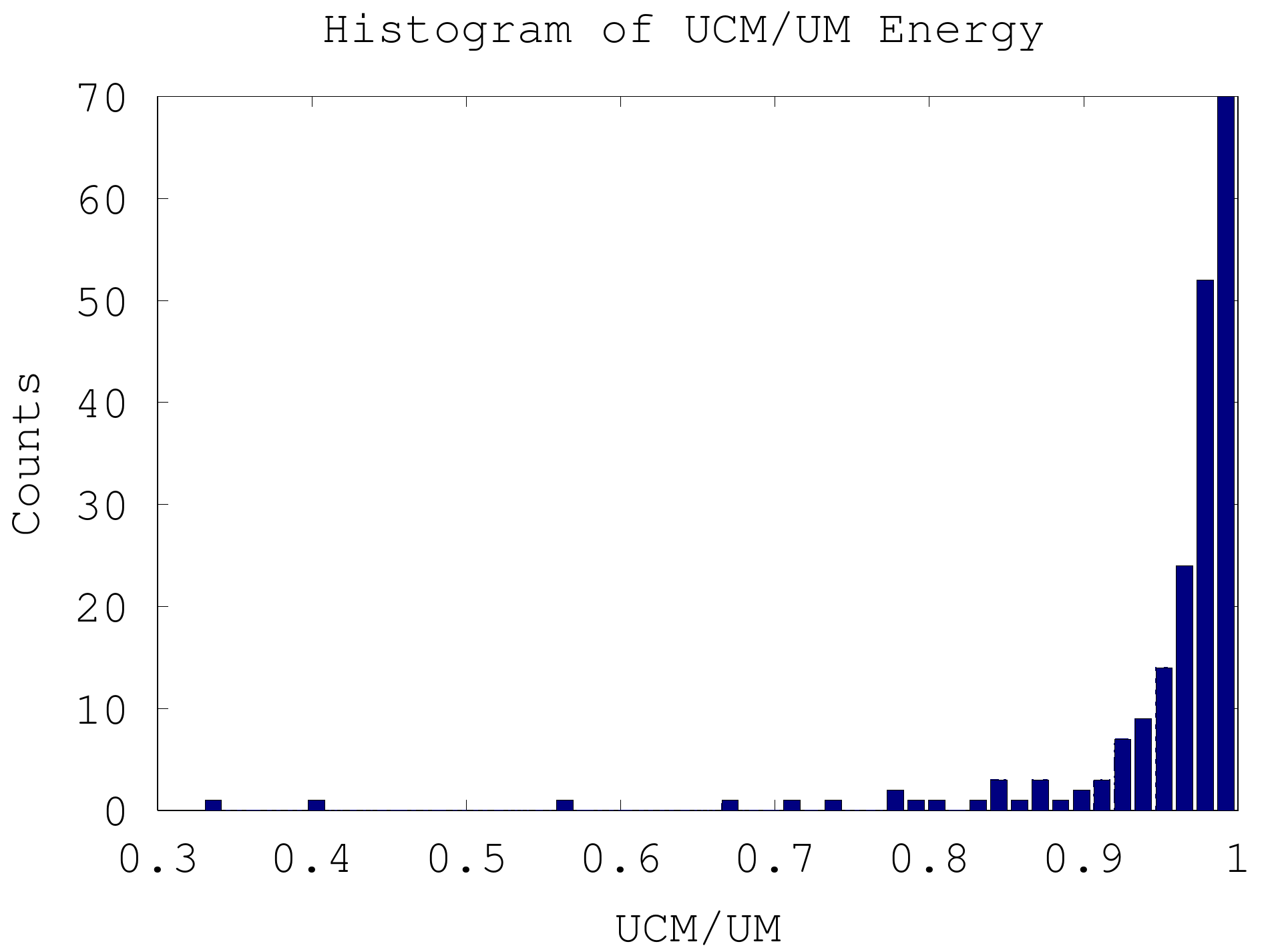}
\caption{
We compare the quality of the ultrametric rounding produced by our ultrametric
rounding (UM) with the baseline ultrametric contour maps algorithm (UCM) in
terms of the ultrametric rounding objective.  We plot a histogram of the ratio
of objective values of UCM and UM.  All ratios were less than $1$ showing that
in no instances did UM produce a worse solution than UCM }
\label{ucmvsum}
\end{figure}

In Fig \ref{ucmvsum} we display a histogram, computed over test image problem
instances, of the cost of UCM solutions relative to those produced by UM
rounding.  A value of 1 indicates equality.  A value of greater than 1
indicates UCM providing lower cost while a value less than 1 indicated UM
providing lower cost.  In no instance did UCM outperform our UM algorithm
though our UM algorithm often outperformed UCM.  

\subsection{Segmentation performance and running time}

\begin{figure*}
\centering     
\subfigure[Precision Recall Curve after 5 seconds]{\label{quantprb:a}\includegraphics[width=65mm,clip,trim=5mm 60mm 5mm 60mm]{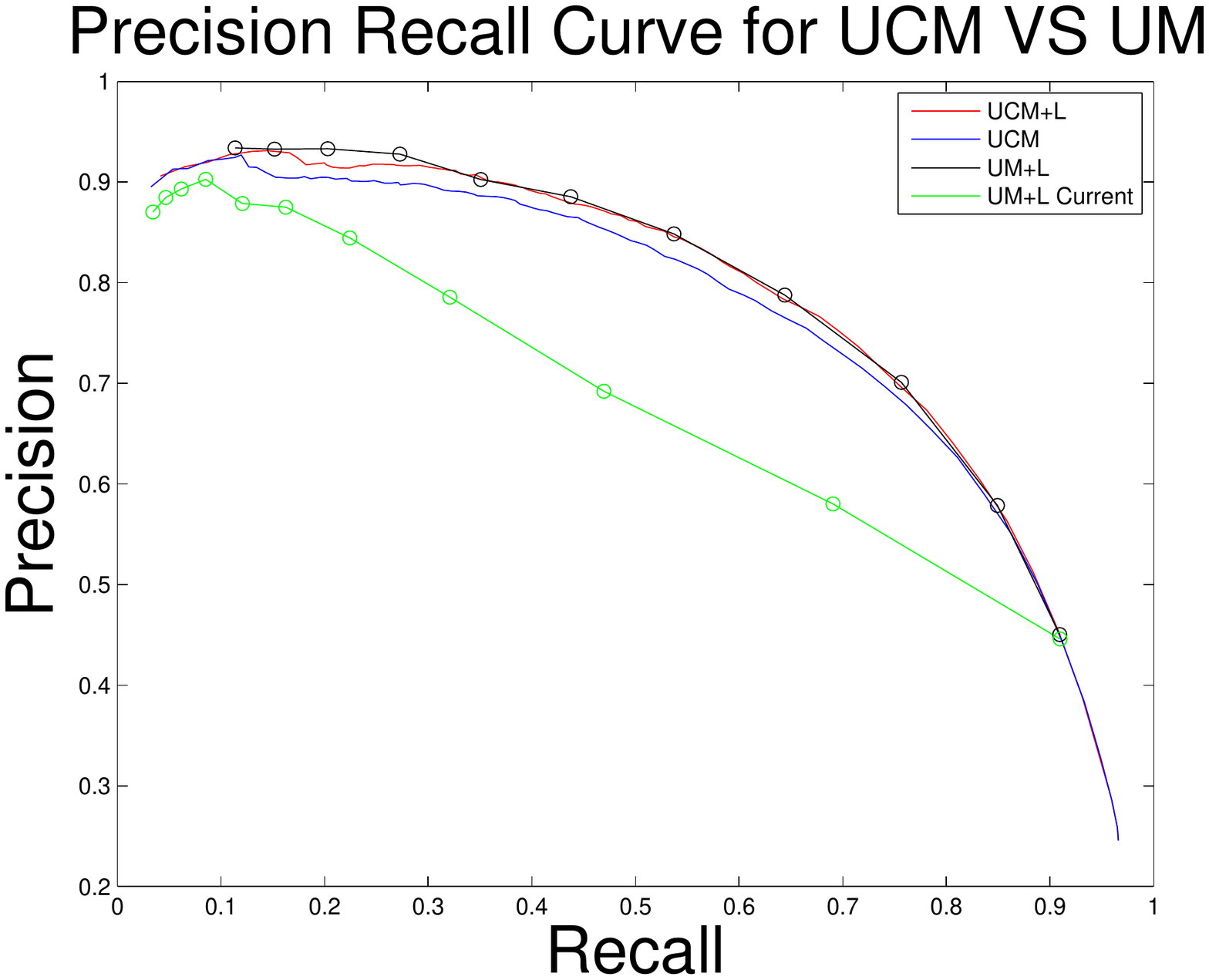}}
\subfigure[Precision Recall Curve after 10 seconds]{\label{quantprb:b}\includegraphics[width=65mm,clip,trim=5mm 60mm 5mm 60mm]{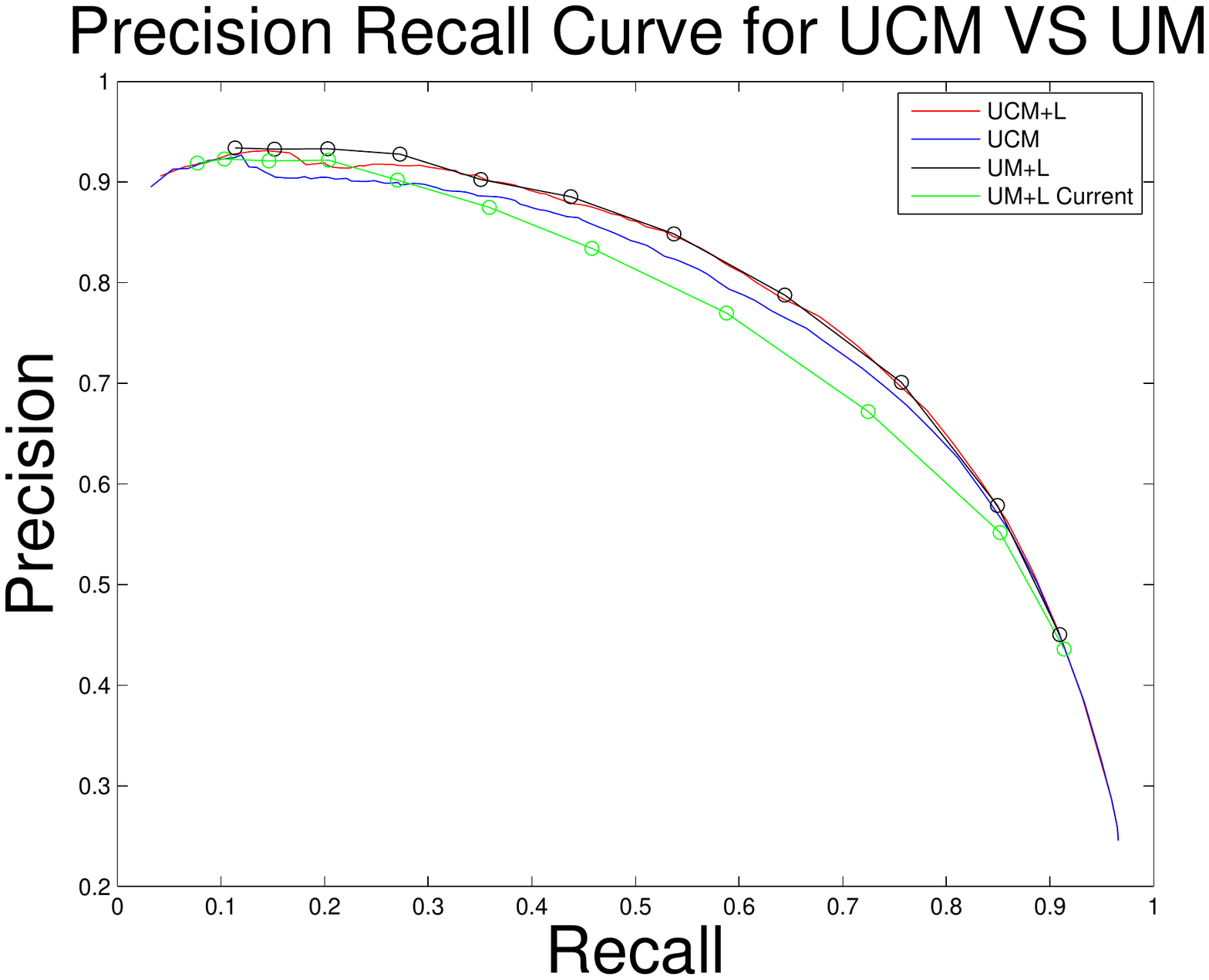}}
\subfigure[Precision Recall Curve after 15 seconds]{\label{quantprb:c}\includegraphics[width=65mm,clip,trim=5mm 60mm 5mm 60mm]{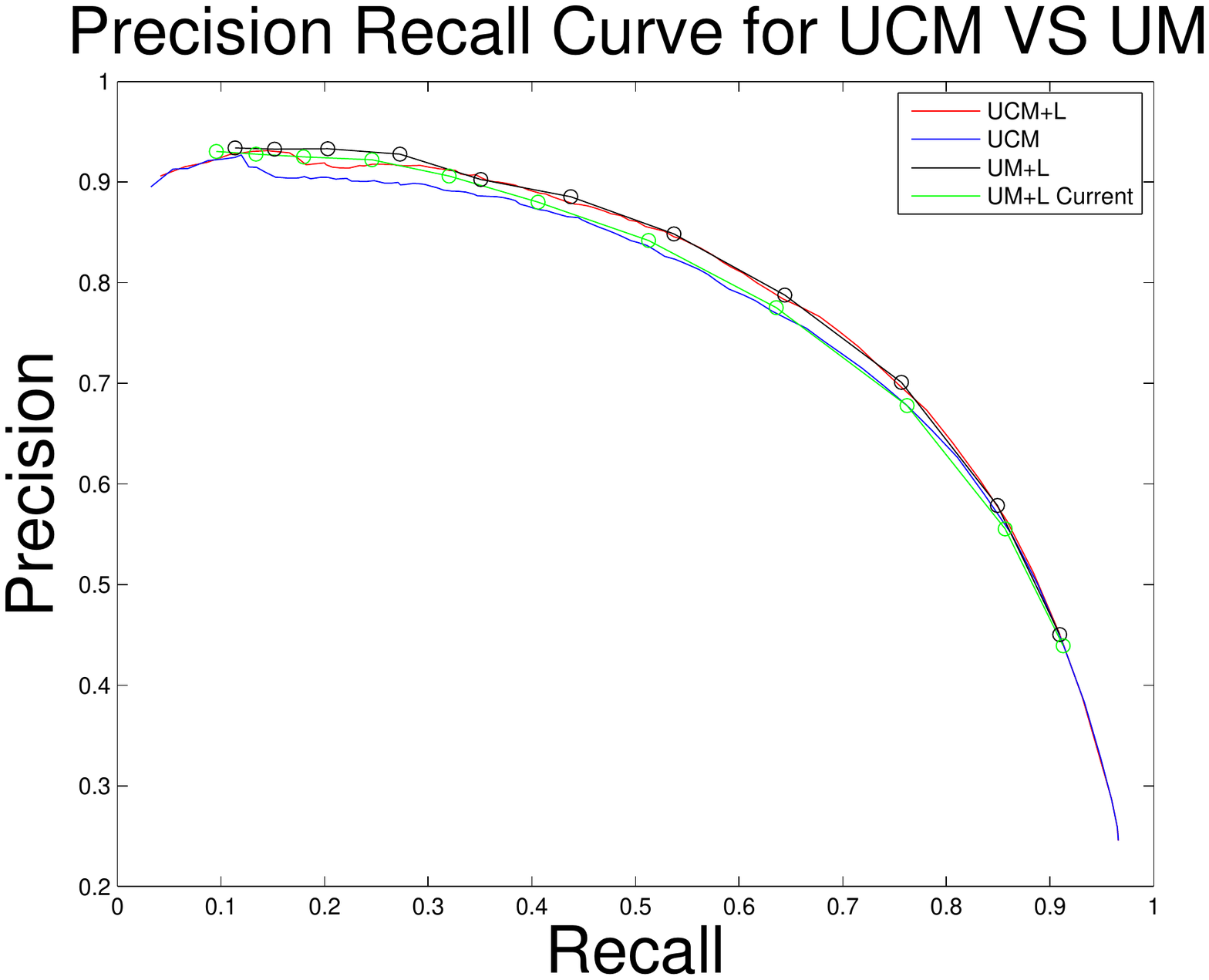}}
\subfigure[Precision Recall Curve after 30 seconds]{\label{quantprb:d}\includegraphics[width=65mm,clip,trim=5mm 60mm 5mm 60mm]{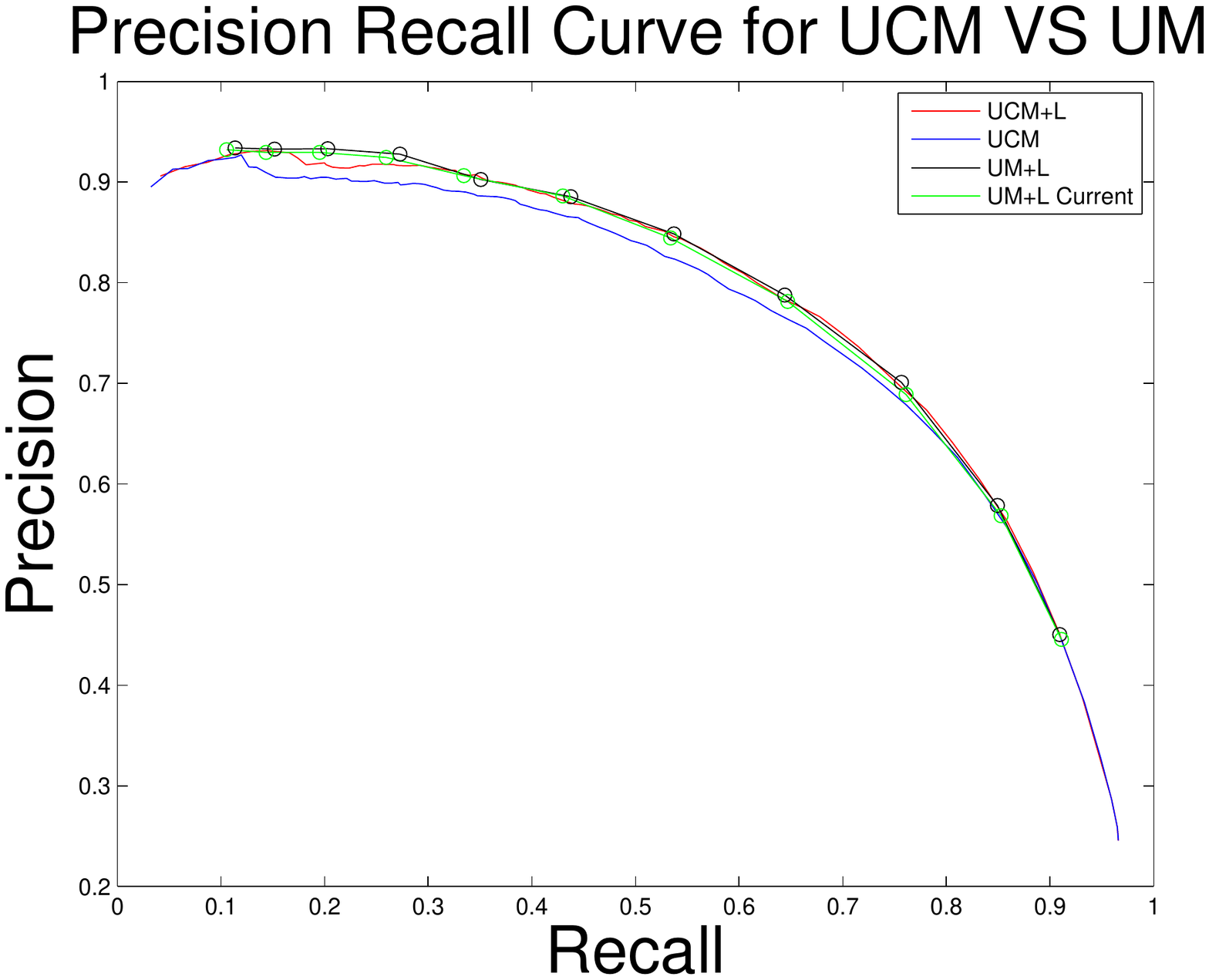}}
\caption{Anytime performance: We show the precision-recall curve of for
segmentations derived from the lowest-cost solution decoded at a particular
amount of execution time (green curves), stopping at T=5,10,15 and 30 seconds
respectively.  We conclude that high-tolerance numerical convergence is not
necessary to achieve good quality segmentations.  For comparison, we plot the
UCM with and without length weighting in red and blue respectively and the UM
results after all problems terminate in black.
}
\label{quantprb}
\end{figure*}

\begin{figure}
	\centering
		\includegraphics[width=0.6\linewidth,clip,trim=5mm 60mm 5mm 60mm]{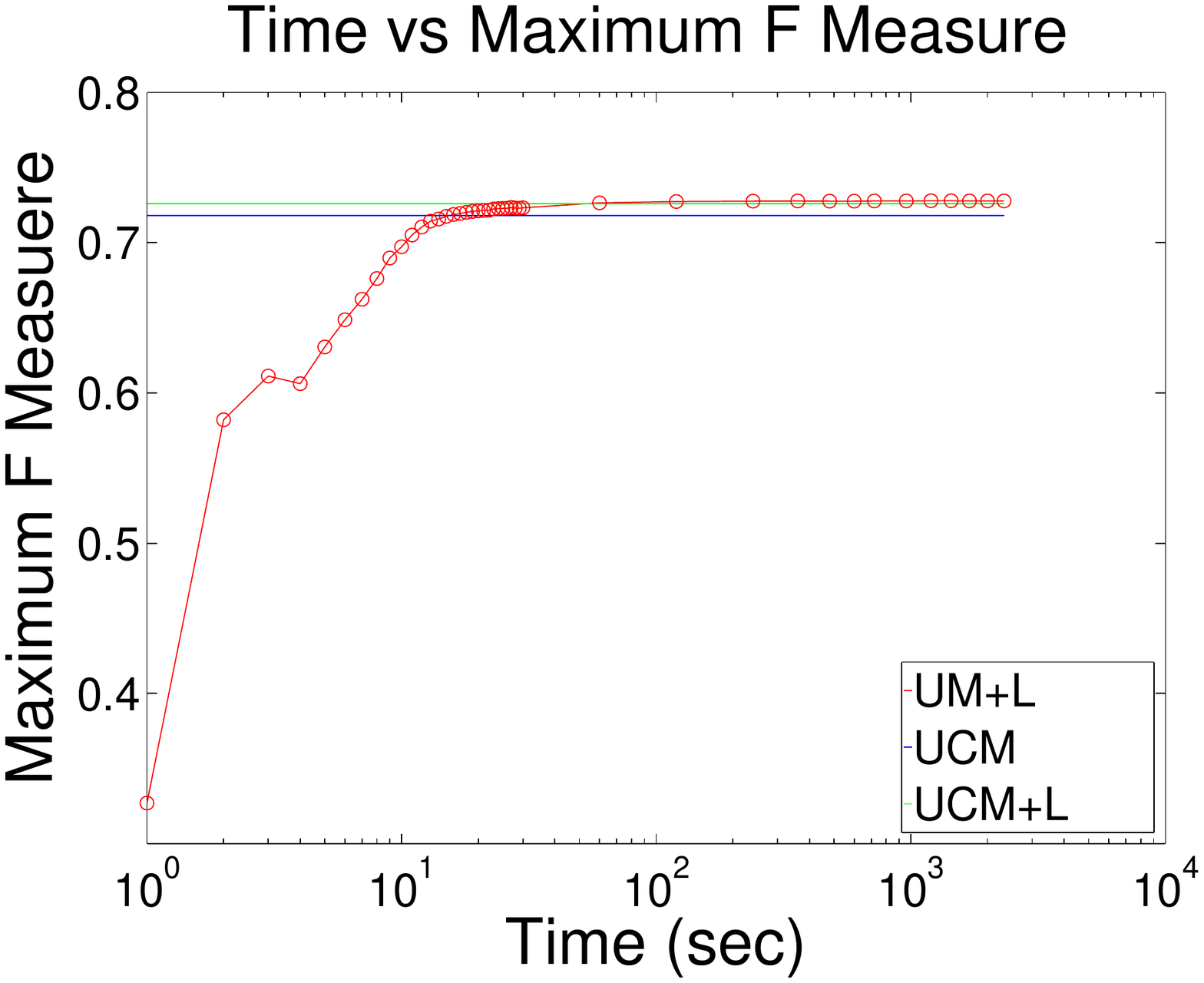}
\caption{
Anytime performance: We plot the maximum F-measure on the BSDS benchmark as a
function of run-time.  Clock time includes lower-bound optimization and
upper-bound decoding after each iteration.  We also include the maximum
F-measure produced by UCM with and without length weighting.   The final
F-measures achieved by UCM, UCM+L and UM are 0.728, 0.726, 0.718 respectively. }
\label{f_meas_plot}
\end{figure}

While our cutting-plane approach is slower than agglomerative clustering, it 
is not necessary to wait for convergence in order to produce high quality results.
We found that while the upper and lower bounds decrease as a function of
time the clustering performance as measured by precision-recall stabilized
is often nearly optimal after only ten seconds and is very stable after that.
We show PR curves at several time points in Fig \ref{quantprb}.  In Fig
\ref{f_meas_plot} we shows a plot of the maximum f-measure of UM rounding as a
function of time relative to the final values of UCM with and without length
weighting.

\subsection{Importance of enforcing hierarchical constraints}
\label{endExper}

\begin{figure*}
\begin{tabular}{m{5mm}m{28mm}m{28mm}m{28mm}m{28mm}}
UM&\includegraphics[width=28mm]{image_example_bad/100007_heir/100007_1_HPC} & 
\includegraphics[width=28mm]{image_example_bad/100007_heir/100007_7_HPC} & 
\includegraphics[width=28mm]{image_example_bad/100007_heir/100007_9_HPC} & 
\includegraphics[width=28mm]{image_example_bad/100007_heir/100007_10_HPC} \\
CC&\includegraphics[width=28mm]{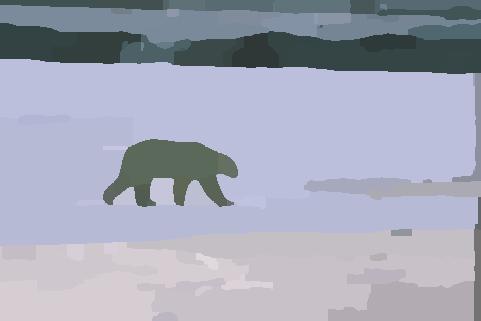} &
\includegraphics[width=28mm]{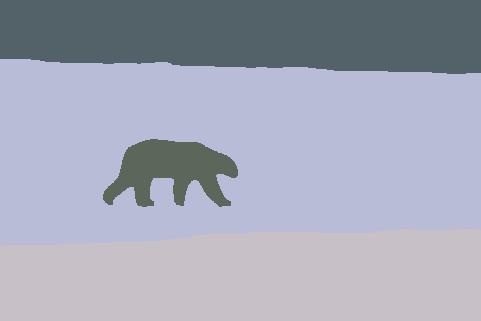} &
\includegraphics[width=28mm]{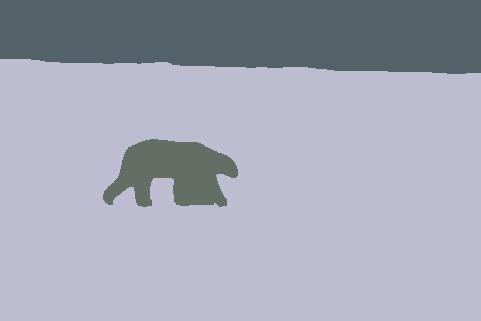} &
\includegraphics[width=28mm]{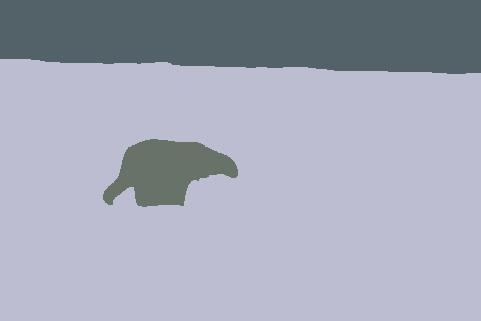} \\
UM&\includegraphics[width=28mm]{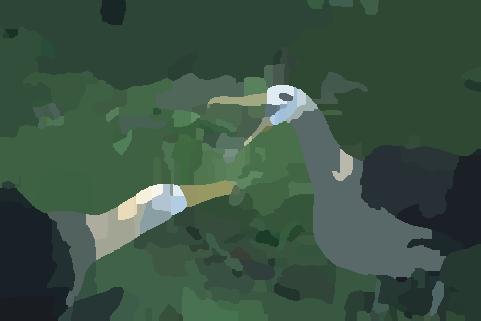} &
\includegraphics[width=28mm]{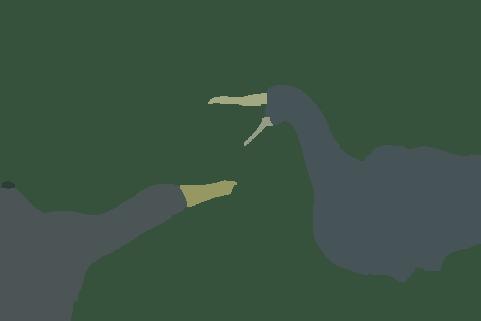} &
\includegraphics[width=28mm]{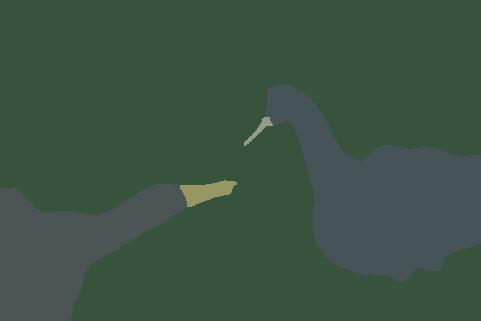} & 
\includegraphics[width=28mm]{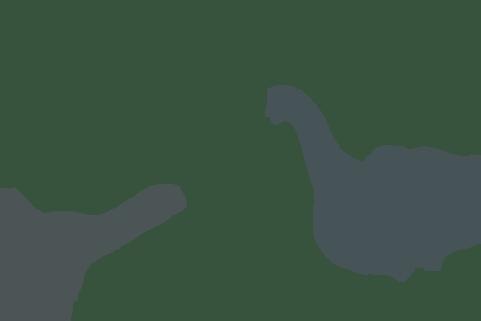} \\ 
CC&\includegraphics[width=28mm]{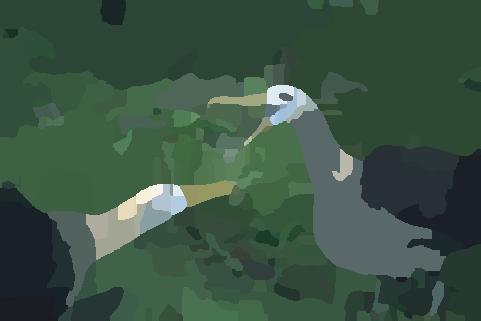} & 
\includegraphics[width=28mm]{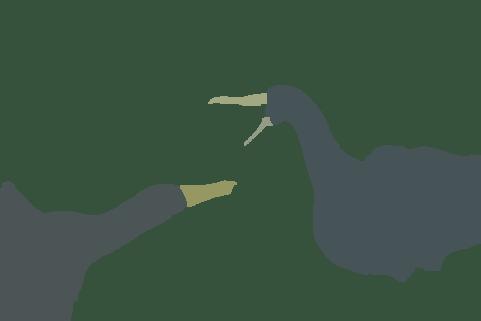} &
\includegraphics[width=28mm]{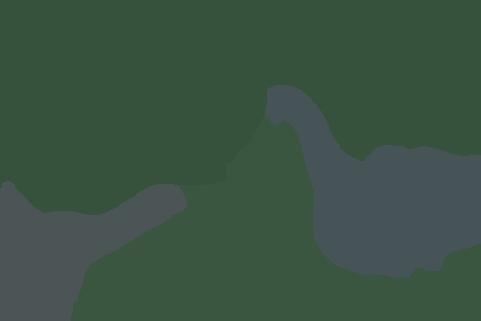} & 
\includegraphics[width=28mm]{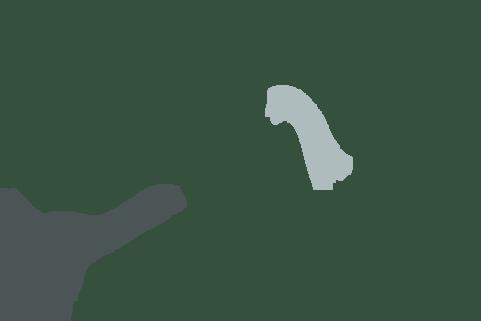} \\
UM&\includegraphics[width=28mm]{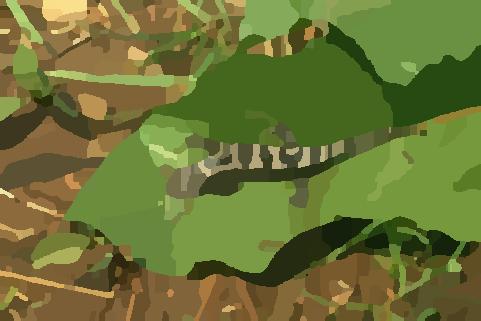} &
\includegraphics[width=28mm]{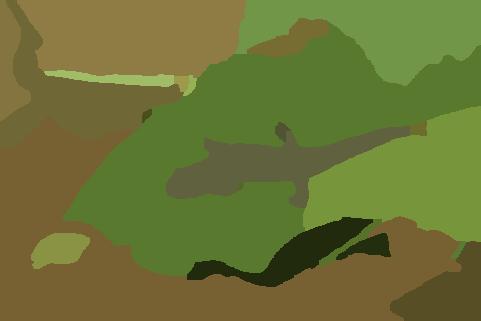} &
\includegraphics[width=28mm]{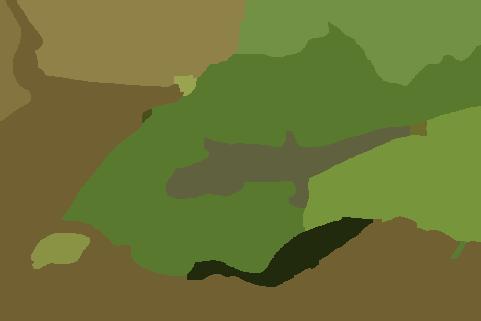} &
\includegraphics[width=28mm]{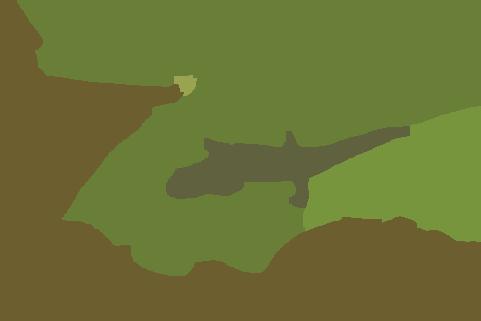} \\
CC&\includegraphics[width=28mm]{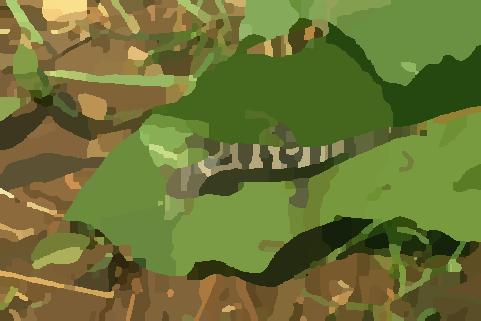} &
\includegraphics[width=28mm]{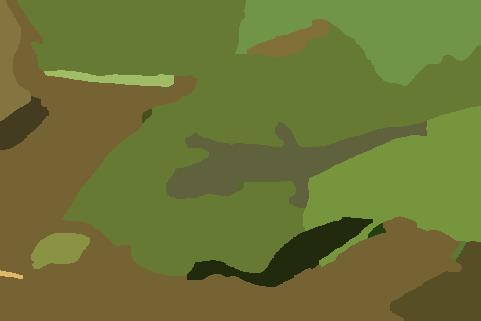} &
\includegraphics[width=28mm]{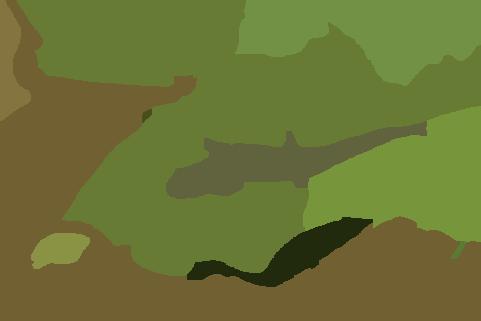} &
\includegraphics[width=28mm]{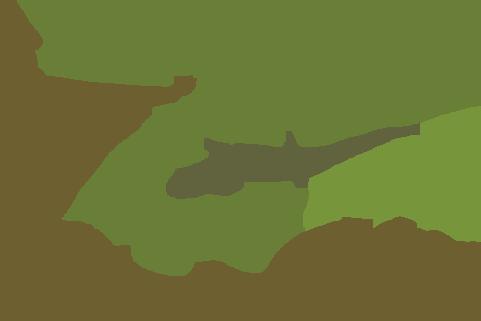} \\
\end{tabular}
\caption{Examples where hierarchically nested segmentations give more
semantically meaningful groupings of the image.  The proposed ultrametric
rounding (UM) enforces consistency across levels while performing independent
correlation clustering (CC) at each threshold does not guarantee a hierarchical
segmentation (c.f. first image). In the second image, hierarchical segmentation
(UM) preserves semantic parts of the two birds while merging the background
regions.  In the third image, CC merges the background clutter into foreground
leaf region at a very low threshold due to a single weak edge.
}
\label{greednogood}
\end{figure*}

Although independently finding multicuts at different thresholds often produces
hierarchical clusterings, this is by no means guaranteed.  We ran Algorithm 1
while enforcing that $\omega^l_e=0$ $\forall [e\in E,l]$.  This allows the
multicut problem for each layer to be solved independently as if the others did
not exist.  To solve these multicut problem instances we used the
solver of \cite{PlanarCC}.  In our data set of 200 images and 11 layers per
problem results in 2200 total multicut instances. The less
constrained single-layer solver produced a lower or equal cost multicut
compared to the hierarchical solver in 99.77 percent of problem instances.  In
Fig \ref{greednogood} we show examples of hierarchy constraints being violated
severely on multiple images when solving with $\omega$ forced to zero.
Introduction of the hierarchy constraint fixes such errors.  

\section{Conclusion}

We have introduced a new method for ultrametric rounding on planar graphs that
is applicable to hierarchical image segmentation.  Our contribution is a dual
cutting plane approach that exploits the introduction of novel slack terms that
allow for representing a much larger space of solutions with relatively few
cutting planes. This yields an efficient algorithm that provides rigorous
bounds on the quality the resulting solution.  We empirically observe that our
algorithm rapidly produces compelling image segmentations along with lower- and
upper-bounds that are nearly tight on the benchmark BSDS test data set.   

\bibliographystyle{plain}
\bibliography{my_bib}

\appendix


\section{Expanded multicut objective and the cycle inequalities}
\label{app1cyc}

In this appendix we show that for planar graphs, solving the expanded multicut
optimization produces solutions that satisfy the cycle inequalities and have
equivalent cost when truncated to lie in the unit hypercube. This establishes
an equivalence between the expanded multicut optimization
\begin{equation}
\min_{\substack{\gamma \geq 0\\ \beta \geq 0}}\theta \cdot {\hat Z}\gamma-\theta^{-} \cdot  \beta \quad \quad 
s.t.\; \; {\hat Z}\gamma -\beta \leq 1 
\end{equation}
and the cycle polytope relaxation
\begin{equation}
\min_{X \in \cycpoly} \theta \cdot X
\end{equation}
for the case of planar graphs.

\subsection{Multicut cone and Cycle cone}
Recall that $\cut$ and $\mcut$ denote the set of binary indicator vectors that
represent valid two-way cuts and multicuts respectively for a specified graph
$G$. We denote the conic hulls of these sets by
\begin{align}
&\cutcone = \left\{\sum_i X^i \gamma_i : \gamma_i \geq 0, X^i \in \cut \right\} \\
&\mcutcone = \left\{\sum_i X^i \gamma_i : \gamma_i \geq 0, X^i \in \mcut \right\}\\
\end{align}
Finally, we denote the cone of positive vectors satisfying the cycle
inequalities by:
\begin{align}
\cyccone = \left\{X \geq 0, \sum_{e \in c -\hat{e}} X_e \geq X_{\hat{e}}, \forall \! c \in C, \hat{e}\in c \right\} 
\end{align}
We now state a two basic results concerning these cones.

\noindent{\bf Proposition 1:} $\mcutcone=\cutcone$\\
Every cut indicator is a multicut indicator, hence $\cutcone \subset
\mcutcone$.  On the other hand, any multicut $X \in \mcut$ can be written as a
conic combination of cuts that isolate each connected component with weight
$\frac{1}{2}$ so that $X = \frac{1}{2} \sum_i Z^i$ with $Z^i \in \cut$ so
$\mcut \subset \cutcone$ and hence $\mcutcone \subset \cutcone$.


\noindent{\bf Proposition 2:} If $G$ is planar, $\cutcone = \cyccone$\\
A stronger version of this result due to \cite{Bar3} states that for a graph
$G$ containing no $K_5$ minor, the set of cycle inequalities over chordless
circuits is sufficient to specify the facets of the cut polytope for $G$.  See
\cite{deza1997geometry} (p. 434) for a detailed discussion.

\subsection{\bf The projected solution $\min(1,Z\gamma)$ satisfies the cycle
inequalities}
As a result of the basic properties of the cut cone, for any $\gamma \geq 0$,
we have $Z\gamma \in \cyccone$ for planar graphs. 
Let $X = \min(1,Z\gamma)$ be a solution to the expanded multicut objective and
$(Z\gamma)_e$ denote the value for a particular edge $e$.  It must then be that
$X \in \cyccone$ since:
\begin{align}
\sum_{e \in c - \hat{e}}\min(1,(Z\gamma)_e) & \geq
\min(1,\sum_{e \in c - \hat{e}} (Z\gamma)_e) \\ 
& \geq 
\min(1, (Z\gamma)_{\hat{e}})
\quad \forall c \in C,\hat{e} \in c
\end{align}
The first inequality arises from pulling the min outside the sum.
The second inequality holds since $Z\gamma \in \cyccone$

\subsection{The projected solution \texorpdfstring{$\min(1,Z\gamma)$}{}
achieves an objective cost no greater than that of \texorpdfstring{$Z\gamma$}{}}
\label{cc3}
We now demonstrate that the fractional multicut $X = \min(1,Z\gamma)$ given by 
projecting the solution $Z \gamma$ yields a solution with an equal or smaller
objective value.

Recall that $\beta$ is a positive slack variable that allows corresponding edge
indicators to take on a value greater than $1$.
\begin{align}
Z\gamma - \beta \leq 1
\end{align}
Since the objective is non-decreasing in $\beta$, for a given setting of
$\gamma$ an optimal setting of the slack variables is given by:
\begin{align}
\beta^*=\max(0,Z\gamma-1)
\end{align}

We split the objective into positive and negative edges and write:
\begin{align}
\theta \cdot Z \gamma - \theta^- \cdot \beta  &= \theta^+ \cdot Z \gamma + \theta^- \cdot Z \gamma - \theta^- \cdot \beta \\
&= \theta^+ \cdot Z \gamma + \theta^- \cdot \min(1,Z \gamma) \\
&\geq \theta^+ \cdot \min(1,Z \gamma) + \theta^- \cdot \min(1,Z \gamma) \\
&=  \theta \cdot \min(1,Z \gamma) \\
&= \theta \cdot X
\end{align}
which establishes that projecting $Z\gamma$ onto the unit cube yields a 
fractional multicut solution that does not increase the objective.


\section{Expanded ultrametric objective and fractional ultrametrics}
\label{ProjectUltraha}
Recall the set of fractional ultrametrics is defined as follows 
\begin{align}
\label{ver0}
\Omega_L = \left\{ \{ X^1,X^2,\ldots X^L\} : X^l \in \cycpoly, X^l \geq X^{l+1} \; \forall l \right\}
\end{align}
In analogy with the previous appendix, we show the equivalence
of the expanded ultrametric rounding problem:
\begin{align}
\min_{\substack{\gamma \geq 0 \\ \beta \geq 0 \\ \alpha \geq 0}}
  \sum^L_{l=1} & \theta^l \cdot Z \gamma^l + \sum^L_{l=1} -\theta^{-l} \cdot \beta^l + \sum^{L-1}_{l=1}  \theta^{+l} \cdot  \alpha^l &\\
s.t. \; \; \nonumber & Z\gamma^{l+1}+\alpha^{l+1} \leq Z \gamma^{l}+\alpha^l \quad \forall l <L\\
& Z\gamma^l-\beta^l \leq 1\quad \forall l
\end{align}
with the relaxed problem:
\begin{align}
\min_{\substack{\mathcal{X} \in \Omega_L}} \sum_{l=1}^L  \theta^l \cdot X^l 
\end{align}

Given an optimal solution to the expanded ultrametric rounding problem
specified by $(\gamma,\alpha,\beta)$, we produce a fractional ultrametric $H$
by the projection operation:
\begin{align}
\label{defH1}
H^l&=\min(1,\max_{m\geq l}(Z\gamma^m)) =\max(H^{l+1},\min(1,(Z\gamma^l)))
\end{align}
We show that the resulting projection $H$ yields a valid fractional
ultrametric $H \in \Omega_L$ whose cost is no greater than the cost of the
corresponding solution to the expanded objective. 

\subsection{Projecting expanded solutions into $\Omega_L$}
By construction, $H$ satisfies the hierarchical constraint $H^l \geq H^{l+1}$.
We show that $H^l \in \cycpoly$ by induction.  In the previous appendix, we
established that $H^L = \min(1,Z\gamma^L) \in \cycpoly$.  Observe that each
$H^l$ for $l<L$ is the coordinate-wise max of $H^{l+1}$ and
$\min(1,Z\gamma^l)$, both of which are in $\cycpoly$ so we only need show that
$\cycpoly$ is closed under coordinate-wise maximum. 

Let $X^1$ and $X^2$ be two elements of $\cycpoly$ and $X^3 = \max(X^1,X^2)$.
We have $\forall c \in C,\hat{e} \in c$
\begin{align}
\sum_{e \in c - \hat{e}} X^3_e  &= \sum_{e \in c - \hat{e}} \max(X^1_e, X^2_e) \\
&\geq \max( \sum_{e \in c - \hat{e}} X^1_e , \sum_{e \in c - \hat{e}} X^2_e ) \\
&\geq \max(X^1_{\hat{e}}, X^2_{\hat{e}}) = X^3_{\hat{e}}\\
\end{align}
where the first inequality arises from pulling the $\max$ outside the sum
and the second because $X^1$ and $X^2$ each satisfy the cycle inequality.
Hence $X^3 \in \cycpoly$.

\subsection{The cost of \texorpdfstring{$H$}{} is no greater than that of \texorpdfstring{$\{\gamma,\alpha,\beta \}$}{}}
Fixing an optimal solution to the expanded ultrametric problem specified by
$\gamma$ we first note that the optimal values of $\beta$ and $\alpha$ are
given by:
\begin{align}
\beta^{l}&=\max(0,Z\gamma^l-1) \\
\alpha^{l}&= \max_{m \geq l} (Z\gamma^m-Z\gamma^l) 
\end{align}
The formula for $\alpha$ can be developed by starting from layer $L$ and
working down, setting $\alpha$ to the smallest possible value needed to 
satisfy the inter-layer constraints for a given $\gamma$.
\begin{align}
\nonumber \alpha^{L} &= 0 \\
\nonumber \alpha^{L-1} &= \max(0,Z\gamma^L - Z \gamma^{L-1})\\
\nonumber \alpha^{L-2} &= \max(0,Z\gamma^{L} - Z \gamma^{L-2},Z\gamma^{L-1} - Z \gamma^{L-2})\\
\ldots
\end{align}
Since the objective is non-decreasing in $\alpha$ and $\beta$, these values are
the smallest values for which the constraints are satisfied.

Plugging in the settings of the slack variables for each layer $l$ we have:
\begin{align}
\nonumber & \theta^l \cdot Z \gamma^l - \theta^{-l} \cdot \beta^l + \theta^{+l} \cdot  \alpha^l \\
\nonumber &\quad = (\theta^{+l}+\theta^{-l}) \cdot Z \gamma^l - \theta^{-l} \cdot \max(0,Z \gamma^l-1) + \theta^{+l} \cdot \max_{m \geq l}(Z\gamma^m-Z\gamma^l) \\ 
\nonumber &\quad = \theta^{+l} \cdot (Z \gamma^l + \max_{m \geq l}(Z\gamma^m-Z\gamma^l)) + \theta^{-l} \cdot (Z \gamma^l - \max(0,Z \gamma^l-1)) \\
\nonumber &\quad = \theta^{+l} \cdot \max_{m \geq l} Z\gamma^m + \theta^{-l} \cdot \min(1,Z \gamma^l) \\
\nonumber &\quad \geq \theta^{+l} \cdot \min(1,\max_{m \geq l} Z\gamma^m) + \theta^{-l} \cdot \min(1,Z \gamma^l) \\
\nonumber &\quad \geq \theta^{+l} \cdot \min(1,\max_{m \geq l} Z\gamma^m) + \theta^{-l} \cdot \min(1,\max_{m \geq l} Z \gamma^m) \\
\nonumber &\quad = \theta^l \cdot H^l
\end{align}
where the second inequality holds because the $\max$ introduced is multiplied by a negative weight.
Since projection can only remain the same or decrease the cost of each layer,
the total objective must also be no greater than the expanded solution:
\[
\sum_l \theta^l \cdot Z \gamma^l - \theta^{-l} \cdot \beta^l + \theta^{+l} \cdot  \alpha^l \geq \sum_l \theta^l \cdot H^l
\]


\section{Derivation of Dual Problem}
\label{dualDerv}
Here we give a derivation of the dual objective over the expanded ultrametric
cut cone which we utilize to provide an efficient column generation approach
based on perfect matching.

We introduce two sets of  Lagrange multipliers $\{ \omega^1 \ldots \omega^{L-1} \}$
and $\{ \lambda^1 \ldots \lambda^L \}$ corresponding to the positivity
constraints in Eq \ref{primalultraz}. 
\begin{align}
\min_{\substack{\gamma \geq 0 \\ \beta \geq 0 \\ \alpha \geq 0} }\max_{\omega \geq 0, \lambda \geq 0 }&\sum_{l=1}^L \theta^l Z \cdot \gamma^l -\sum_{l=1}^L \theta^{-l} \beta^l +\sum_{l=1}^{L-1}\theta^{+l}\alpha^{l} \\ 
&\nonumber + \sum_{l=1}^{L-1} \omega^l(Z \cdot \gamma^{l+1} +\alpha^{l+1}-Z\gamma^l -\alpha^l) \\
&\nonumber +\sum_{l=1}^L \lambda^l( Z \cdot \gamma^l-1-\beta^l)
\end{align}

For notational convenience, we set $\alpha^L=0$ and $\omega^0=0$.  We reorder the 
terms of the Lagrangian in terms of summations over the primal variable indices.
\begin{align}
\label{finalDualr1}
\min_{\substack{\gamma \geq 0 \\ \beta \geq 0 \\ \alpha \geq 0} }\max_{\omega \geq 0, \lambda \geq 0 }
& \sum_{l=1}^L -\lambda^l 1 +\sum_{l=1}^L (-\theta^{-l}-\lambda^l )\beta^l \\
&\nonumber+\sum_{l=1}^L (\theta^{+l}+\omega^{l-1}-\omega^l)\alpha^l +\sum_{l=1}^L (\theta^l+\lambda^l +\omega^{l-1}-\omega^{l}) \cdot Z\gamma^l
\end{align}
Each primal variable yields a positivity constraint in the dual.
\begin{align}
\label{finalDualr2}
\max_{\omega \geq 0, \lambda \geq 0 } & \sum_{l=1}^L -\lambda^l 1 & \\
s.t.\; \; 
&\nonumber (-\theta^{-l}-\lambda^l ) \geq 0 \quad & \forall l\\
&\nonumber (\theta^{+l}-\omega^l+\omega^{l-1}) \geq 0 \quad & \forall l\\
&\nonumber (\theta^l+\lambda^l +\omega^{l-1}-\omega^{l}) \cdot Z \geq 0 \quad &\forall l
\end{align}

This dual LP can be interpreted as finding modification of the original edge
weights $\theta^l$ so that every possible cut of each resulting graph has
non-negative weight.  Observe that the
introduction of the two slack terms $\alpha$ and $\beta$ in the primal problem
(Eq \ref{primalultraz}) results in bounds on the Lagrange multipliers $\lambda$
and $\omega$ in the dual problem  in Eq \ref{finalDualr2}. The constraint
$(-\theta^{-l}-\lambda^l )\geq 0$ is a result of the introduction of $\beta^l$.
The constraint $\omega^{l-1}-\omega^l \leq \theta^{+l}$ is a result of the
introduction of $\alpha^l$.  In practice these bounds turn out to be essential
for efficient optimization and are a key contribution of this paper. 

It is also informative to make the substitution $\mu^l = \omega^l -
\omega^{l-1}$ which yields a slightly more symmetric formulation
\begin{align}
\label{dualprobalt}
\max & \sum_{l=1}^L -\lambda^l 1 & \\
s.t.\; \; 
&\nonumber 0 \leq \lambda^l \leq -\theta^{-l} & \forall l \\
&0 \leq \sum_{m=1}^l \mu^m & \forall l \\
&\nonumber  \mu^l \leq \theta^{+l} & \forall l\\
&\nonumber (\theta^l+\lambda^l -\mu^{l})\cdot Z\geq 0 & \forall l 
\end{align}


\section{Producing a genuine lower bound on the optimal integer solution}
\label{genuinelow}
Consider optimizing the Lagrangian over the set of integer solutions
$\mathcal{X} \in \bar{\Omega}_L$. In this case the $\alpha,\beta$
terms disappear.  For a given setting of the remaining multipliers
$\omega,\lambda$ we have a lower bound on the optimal integer solution
given by:
\begin{align}
\nonumber 
L(\omega,\lambda) &= \min_{\mathcal{X} \in \bar{\Omega}_L} \sum_{l=1}^L (  \theta^l\bar{X}^{l} +\omega^l(\bar{X}^{l+1}-\bar{X}^l)+\lambda^l( \bar{X}^{l}-1))\\
\nonumber &=\min_{\mathcal{X} \in \bar{\Omega}_L} \sum_{l=1}^L(\theta^l \bar{X}^{l}+\omega^{l-1}\bar{X}^{l}-\omega^l \bar{X}^{l}+\lambda^l\bar{X}^{l} -\lambda^l1)\\
\nonumber &=\min_{\mathcal{X} \in \bar{\Omega}_L} \sum_{l=1}^L(\theta^l +\omega^{l-1}-\omega^l +\lambda^l )\bar{X}^{l} -\lambda^l1 \\
\nonumber &=\sum_{l=1}^L-\lambda^l1+ \min_{\mathcal{X} \in \bar{\Omega}_L} \sum_{l=1}^L(\theta^l +\omega^{l-1}-\omega^l +\lambda^l )\bar{X}^l \\
\nonumber &\geq \sum_{l=1}^L-\lambda^l1 + \sum_{l=1}^L \min_{X^l \in \mcut}(\theta^l +\omega^{l-1}-\omega^l +\lambda^l )\bar{X}^l  \\
          &\geq  \sum_{l=1}^L-\lambda^l1 + \sum_{l=1}^L\frac{3}{2}\min_{\bar{X}^l \in \cut}(\theta^l +\omega^{l-1}-\omega^l +\lambda^l )\bar{X}^l
\end{align}
where the first inequality arises from dropping the constraints between layers
of the hierarchy and the second inequality holds for planar graphs where the
the optimal multi-cut is bounded below by $\frac{3}{2}$ the value of the optimal
two-way cut (see \cite{PlanarCC}).

\end{document}